\documentclass[useAMS,usenatbib,twocolumn,fleqn]{mnras}
\usepackage[T1]{fontenc}
\usepackage{ae,aecompl}
\usepackage{natbib, graphics, epsfig}

\usepackage{times}
\usepackage{amsmath}
\usepackage{amssymb}
\usepackage{textcomp}
\usepackage{verbatim}

\newcommand{\hmsun}{\,h^{-1}\,{\rm M_\odot}}
\newcommand{\msun}{{\,\rm M_\odot}}
\newcommand{\kms}{\,{\rm km}\,{\rm s}^{-1}}

\newcommand{\pc}{\,{\rm pc}}
\newcommand{\kpc}{\,{\rm kpc}}

\newcommand{\hkpc}{\,h^{-1}\,{\rm kpc}}
\newcommand{\hmpc}{\,h^{-1}\,{\rm Mpc}}
\newcommand{\cpm}{\,{\rm cm}^2\,{\rm g}^{-1}}
\usepackage{color}

\def\jcap{J. Cosmol.  Astropart. Phys.}
\def\aap{A\&A}
\def\apj{ApJ}

\def\apjl{ApJ}
\def\mnras{MNRAS}

\def\aj{AJ}

\def\physrep{Phys. Rep.}
\def\nat{Nature}

\def\apjs{ApJS}
\def\prd{Phys. Rev. D}

\title[Dark matter physics as a possible explanation of the small-scale CDM problems]
      {ETHOS - An Effective Theory of Structure Formation: Dark matter physics as a possible explanation of the small-scale CDM problems}
      \author[M. Vogelsberger et al.] {\parbox{18.5cm}{
	  Mark Vogelsberger$^1$\thanks{e-mail: mvogelsb@mit.edu}, 
          Jes\'us Zavala$^2$\thanks{Marie Curie Fellow},
          Francis-Yan Cyr-Racine$^{3,4}$, 
          Christoph Pfrommer$^5$,
          Torsten Bringmann$^6$, and       
          Kris Sigurdson$^{7,8}$
        }\vspace{0.3cm}\\
        $^1$ Department of Physics, Kavli Institute for Astrophysics and Space Research, Massachusetts Institute of Technology, Cambridge, MA 02139, USA\\	
        $^2$ Dark Cosmology Centre, Niels Bohr Institute, University of Copenhagen\\
        $^3$ Department of Physics, Harvard University, Cambridge, Massachusetts 02138, USA\\
        $^4$ California Institute of Technology, Pasadena, CA 91125, USA\\
        $^5$ Heidelberg Institute for Theoretical Studies, Schloss-Wolfsbrunnenweg 35, D-69118 Heidelberg, Germany\\
        $^6$ Department of Physics, University of Oslo, Box 1048 NO-0316 Oslo, Norway\\
        $^7$ School of Natural Sciences, Institute for Advanced Study, Einstein Drive, Princeton, NJ 08540\\
        $^8$ Department of Physics and Astronomy, University of British Columbia, Vancouver, BC, V6T 1Z1, Canada}

\date{Accepted XXX. Received YYY; in original form ZZZ}

\pubyear{2015}

\begin{document}
\label{firstpage}
\pagerange{\pageref{firstpage}--\pageref{lastpage}}
\maketitle





\begin{abstract}
We present the first simulations within an effective theory of structure
formation (ETHOS), which includes the effect of interactions between dark
matter and dark radiation on the linear initial power spectrum and dark matter
self-interactions during non-linear structure formation. We simulate a Milky
Way-like halo in four different dark matter models and the cold dark
matter case. Our highest resolution simulation has a particle mass of
$2.8\times 10^4\msun$ and a softening length of $72.4\pc$. We demonstrate that
all alternative models have only a negligible impact on large scale structure
formation.  On galactic
scales, however, the models significantly affect the structure and abundance of
subhaloes due to the combined effects of small scale primordial damping in the
power spectrum and late time self-interactions. We derive an analytic mapping from the primordial damping scale in
the power spectrum to the cutoff scale in the halo mass function and the
kinetic decoupling temperature. We demonstrate that certain
models within this extended effective framework that can alleviate the
too-big-to-fail and missing satellite problems simultaneously, and possibly the
core-cusp problem. The primordial power spectrum cutoff of our
models naturally creates a diversity in the circular velocity profiles, which is larger than that found for cold dark matter simulations. We show that the parameter space of models can be constrained by contrasting
model predictions to astrophysical observations. For example, some
models may be challenged by the missing satellite problem if baryonic processes
were to be included and even over-solve the too-big-to-fail problem; thus
ruling them out.
\end{abstract}


\section{Introduction}

Despite its uncertain nature, dark matter (DM) is the key driver of structure
formation in the Universe. The current paradigm,  the so-called Cold DM (CDM) model,
assumes that DM is cold and collisionless~\citep[][]{Blumenthal1984,
Davis1985}.  This model is extremely successful in describing the large scale
structure of the Universe~\citep[e.g.][]{Springel2005, Vogelsberger2014}.
However, significant small-scale discrepancies on galactic and sub-galactic scales remain.
Specifically, (i) the problem of the abundance of dwarf galaxies, well known as
the ``missing satellite (MS) problem'' for galaxies in the Milky Way
(MW)~\citep[][]{Klypin1999,Moore1999}, and more recently pointed out also for
galaxies in the field \citep[][]{Zavala2009,Papastergis2011,Klypin2014}, (ii)
the ``too-big-to-fail (TBTF)
problem''~\citep[][]{Boylan-Kolchin2011,Papastergis2015} and the possibly related diversity of dwarf rotation curves~\citep[][]{Oman2015}, (iii) the core-cusp (CC)
problem for low surface brightness galaxies~\citep[][]{deBlok1997}, and dwarf
galaxies~\citep[e.g. see][]{Oh2011, Walker2011}, and (iv) the
plane of satellites problem~\citep[][]{Pawlowski2013}.  

It is at the moment unclear whether all these challenges can be fully explained
within the CDM framework by invoking the complex baryonic physics in galaxy
formation and evolution.  
However, it is certain that the CDM problems are only firmly established in the discrepancies
between CDM-only simulations and observations.  Some of the baryonic processes that
are known to happen in galaxies might actually be able to resolve these issues.
Core formation, for example, might be driven by sufficiently strong supernova
feedback~\citep[e.g.][]{Navarro1996,Pontzen2012,Sawala2014,Gillet2015,Chan2015,Brook2015}, while gas heating during the
reionisation era might prevent the gas from cooling and forming stars
efficiently in low-mass haloes, making them invisible to current surveys rendering
the MS problem less severe.  
Unfortunately, there is limited
evidence that can confirm if these baryonic processes can be as efficient as they
need to be to become viable solutions. For instance, bursty star formation
histories, occurring at shorter time scales than the characteristic dynamical
time of the galaxy, are required for supernovae feedback to dramatically alter
the inner DM distribution, but such time resolution is still not possible for
the dwarf galaxies that show the CDM problems~\citep[][]{Weisz2014}, although
there is evidence that this process is indeed efficient at larger masses
~\citep[][]{Kauffmann2014}. Furthermore, even optimistic
baryonic solutions to these CDM problems fail to explain the diversity problem of dwarf
galaxy rotation curves as pointed out by~\cite{Oman2015} (a problem already hinted at by \citealt{Kuzio2010} for LSB galaxies, see their Fig. 4).  
On the other hand, the plane of satellites issue, could even be fully consistent with CDM in general and therefore
represent no challenge to CDM~\citep[see for example][]{Cautun2015}. 
Regarding the abundance of dwarf galaxies,  the explanation of the MS problem, which may involve a combination of 
observational incompleteness of current surveys and environmental processes that
suppress star formation in the satellites, might not explain the dearth of field dwarfs.
Here, environmental effects are likely not present -- although there could be globally acting feedback like blazar heating~\citep[][]{Pfrommer2012} -- and surveys based on the
detection of gas, such as ALFALFA, should be sensitive enough to detect most
gas-bearing dwarf galaxies even if their stellar content is low. Yet, galactic winds
driven by supernovae (SNe) explosions in galaxies with low stellar masses might well
be strong enough to expel the gas from their dark matter halos, and 
make these dwarfs invisible to cold gas surveys as well
(e.g. \citealt{Sawala2012}).  As this issue remains inconclusive, theoretical
predictions need to be confronted with an updated census of galaxies in the
local Universe \citep{Klypin2014}. It is fair to state that, at the moment, the TBTF problem, the abundance problem, the CC problem and 
the diversity of dwarf galaxy rotation curves are not convincingly solved within CDM even taking into
account the effects of baryons, albeit different paths towards a baryonic-only solution have been recently proposed, e.g.
\citet{Brook2015,Brook2016,Brook2015b,Chan2015}. 

This motivates us to think about DM physics beyond
simple CDM models, which might have a prominent role in the outstanding
small-scale issues. In the CDM model, DM is assumed to be cold and
collisionless. Modifications of these two basic, so far unproven DM properties,
lead to differences in the nonlinear structure formation process that might
help to solve some of the outstanding small-scale issues of the CDM paradigm.
Since the success of the CDM model needs to be preserved on large scales, one
is mainly looking for small ``perturbations'' around the concordant CDM model.
There are mainly two alternatives studied in the context of solving galaxy
formation problems that relax these hypotheses separately: warm DM (WDM) and
self-interacting DM (SIDM).  

While viable WDM models clearly lead to a suppression of the abundance of
small haloes and a reduction of central halo densities, they do not lead to a
substantial modification of the inner density profile of DM haloes, preserving
the ``universal'' character of the CDM Navarro-Frenk-White profile.  Most importantly $\sim {\rm keV}$ thermal relic WDM does not create cores
on astrophysical scales~\citep[][]{Villaescusa-Navarro2011,Maccio2012}, and current Lyman-$\alpha$
forest measurements on the DM power spectrum tightly constrain thermal relics. For example,
\cite{Viel2013} finds a lower limit of $m_{\rm WDM} \gtrsim 3.3\,{\rm keV}$,
which is too large to provide solutions to the small scale problems of CDM that
require $m_{\rm WDM}\sim2\,{\rm keV}$~\citep[][]{Schneider2014}.
Nevertheless, even modest small scale modifications of the initial power spectrum lead
to effects on the satellite population of haloes, that are 
interesting in the context of the MS problem. Thermal relic WDM is most likely not the right idea
to provide these modifications in a way consistent with current observations and constraints (although see the recent analysis by \cite{Garzilli2015}, which relaxes the \cite{Viel2013} constraints significantly). 
There might exist some so far unexplored viable WDM models~\citep[e.g., sterile neutrinos][]{Boyarsky2009}, but only a few N-body simulations with these models have been done~\citep[e.g.,][]{Lovell2012}.

On the other hand, the internal mass distribution of DM haloes can be modified
substantially in allowed SIDM models, forming density cores through an
effective inwards heat flux that drives haloes towards an isothermal configuration.
If the self-scattering cross section per unit mass is $\sim1\cpm$ at the scale
of dwarf galaxies, SIDM models can solve both the CC and TBTF problems
\citep{Vogelsberger2012,Rocha2013,Zavala2013}. In SIDM, the (sub)halo abundance
can potentially be modified as well if the cross section is large enough, which
was one of the original motivations of the SIDM model ~\citep{Spergel2000}.
However, the cross sections required for this to happen are seemingly too
large, of $\mathcal{O}(10\cpm$) on galactic scales, clearly above the most
up-to-date constraints, which suggest that constant cross sections larger than
$1\cpm$ are ruled out~\citep[][]{Peter2012}. At the level of these bounds,
there are no significant differences between the SIDM and CDM (sub)halo
abundances~\citep[][]{Vogelsberger2012,Rocha2013}. It is important to recall
however, that such conclusions are based on DM-only simulations.  Recent
analytical estimates suggest that the constraints might get weaker once the
effects of baryons are considered~\citep[][]{Kaplinghat2014}. Additionally,
since current constraints are all beyond the scale of massive
ellipticals~\citep[see however][for some model dependent constraints on other
mass scales, which point to a slight velocity dependence of the cross
section]{Kaplinghat2015}, the possibility remains that DM has very large cross
section on smaller (dwarf-size) scales, dropping below current bounds at larger
scales. This velocity-dependent cross section is also the natural outcome of
many particle physics models of
SIDM~\citep[e.g.][]{Ackerman2009,Feng2009,Feng2010,Buckley2010,Loeb2011}.  
On the observational side, \cite{Massey2015} recently found an offset between the inferred position
of the DM halo and the stars of one of the $4$ bright cluster galaxies in the
$10\kpc$ core of Abell $3827$. This offset could be interpreted as potential
evidence for SIDM with a cross-section of $\sim1.5\cpm$ \citep{Sarkar2015}, in
marginal tension with current constraints.  Although such an offset might be
caused by astrophysical effects unrelated to the DM nature, its amplitude appears 
to be extremely rare within the CDM model \citep{Schaller2015}.

We note that the SIDM and WDM models explored in the context of structure
formation are just a subgroup of a broader class of possible alternative DM
models. To mention a couple of less explored models, DM from late decays
\citep[][]{Sigurdson2004,Kaplinghat2005,Borzumati2008}, and extremely light DM
forming condensates \citep{Hu2000,Peebles2000} with particle masses of the
order of $10^{-22}{\rm eV}$. The latter class of models is particularly
interesting since non-QCD axions could be a potential candidate for such light
DM, which would be described by a single coherent wave function~\citep[see][for
some simulation results]{Schive2014}.

We would like to stress that all these different modifications to the CDM
paradigm are also motivated by the fact that the most simple CDM candidates
(Weakly Interacting Massive Particles, WIMPs) have not been discovered despite
decades long efforts~\citep[e.g.,][]{Bertone2005,Bertone2010}.  Furthermore,
there has also been no sign  Supersymmetry at the LHC so far, which has
provided a strong theoretical support for several excellent CDM-WIMP candidates
like the neutralino ~\citep[][]{Jungman1996}.  Based on these null results, the 
time is ripe to think beyond simple CDM models.  The fact that some of these
alternative models can actually also solve astrophysical problems with CDM
should be seen as a motivation to go beyond the purely particle physics
based considerations. In the end there is also no compelling reason for the fact
that the dark sector should be dominated by a single featureless WIMP
particle only. After all, the visible sector is very rich, and this
might also be true for the dark sector. From the perspective of structure
formation, the CDM model is only an effective description that assumes that the
only DM interaction that matters is gravity.  Since such an assumption remains
unverified, it is crucial that we develop a more generic effective framework that
includes a broad range of allowed DM interactions.

SIDM simulations have so far mainly considered the late time impact of DM
collisions, i.e., those simulations started with a CDM transfer function and
implemented some form of scattering (mostly elastic and isotropic, but see \citealt{Medvedev2014, Randall2014} for first approaches going beyond this) of DM particles, which then affects the
nonlinear structure formation process.  Under these conditions, SIDM
simulations have shown that DM collisions can hardly affect the abundance of
dwarf-size haloes without violating current constraints on the cross-section.
This has led to the impression that SIDM by itself is not able, for example, to
solve or alleviate the CDM problem on the abundance of dwarf
galaxies~\citep[][]{Brooks2014}.  
However, most previous SIDM simulations neglect 
the possibility of modifying the primordial linear power spectrum. Although DM
self-collisions are not strong enough to impact the power spectrum on galactic
scales, additional interactions with relativistic particles (e.g. photons or
dark radiation) in the early Universe can suppress the formation of dwarf-scale
haloes \citep[][]{Boehm2002, Buckley2014, Boehm2014}.  
This can in fact be naturally accommodated in SIDM models, for exactly the range of parameters that address the TBTF and the CC problem, by allowing the vector field that mediates DM self-interactions to also couple to (sterile) neutrinos or any other form of dark radiation (DR) \citep[][]{vandenAarssen2012, Bringmann2014, Dasgupta2014}. The result is a much later kinetic decoupling of DM than in the standard case~\citep[][]{Bringmann2009}, leading to cutoff scales of the primordial linear power spectrum that are on mass scales of order of dwarf galaxies and hence alleviate the overabundance problem of these (sub)halos in CDM cosmologies~\citep[][]{vandenAarssen2012, Boehm2014, Schewtschenko2014, Buckley2014}.
In general, this suppression of the
initial power spectrum has a similar impact in structure formation as that seen
in vastly studied WDM simulations, but with more complex features that create
richer possibilities.  Depending on the detailed particle physics model, the
power spectrum can have some non-trivial features like dark acoustic
oscillations (DAOs) and a Silk damping tail \citep[e.g.][]{Buckley2014}, which
are a general phenomena of e.g. atomic dark matter
models~\citep[e.g.][]{Cyr-Racine2013}.  Such features are not present in thermal WDM
where the power spectrum suppression is driven by the free streaming of DM
particles mainly resulting in a pure exponential cutoff.  It is however important to mention that 
$N$-body simulations with DAO-like features in the power spectrum have been explored in the past in
the context of different particle physics models, e.g. the consequences for the abundance of (sub)haloes in models 
with long-lived charged massive particles were explored in \citealt{Kamada2013}.

Virtually all previous SIDM simulations have neglected the combined effect of
modifications to the initial power spectrum due to DM interactions with
relativistic particles and DM collisions due to DM self-scattering. There is no
single cosmological simulation that considers a SIDM model,
where the initial power spectrum and self-interactions are consistently chosen.
A first attempt in this direction was made very recently by
\citet{Buckley2014}, but this work did not take into account the proper
velocity-dependence of the scattering cross-section, assuming instead a
constant cross section. 
We present here the first $N-$body simulations where
viable particle physics models are used as an input to 
compute the initial power spectrum (strongly modified by DM-DR
interactions) and the velocity dependent DM-DM cross section.  Our models have
therefore three main features that affect structure formation: (i) a
suppression of primordial structure due to a Silk-like damping mechanism, (ii)
an imprint of DAOs in the initial matter power spectrum (driven by
DM-DR interactions), and (iii) a modification of DM halo properties
due to the self-interaction of DM in the nonlinear regime of structure
formation. We will demonstrate that such features have interesting effects on
the abundance and the internal structure of haloes and subhaloes, and
thereby successfully address most of the current small-scale issues of CDM.

\begin{figure*}
\centering
\includegraphics[width=0.475\textwidth]{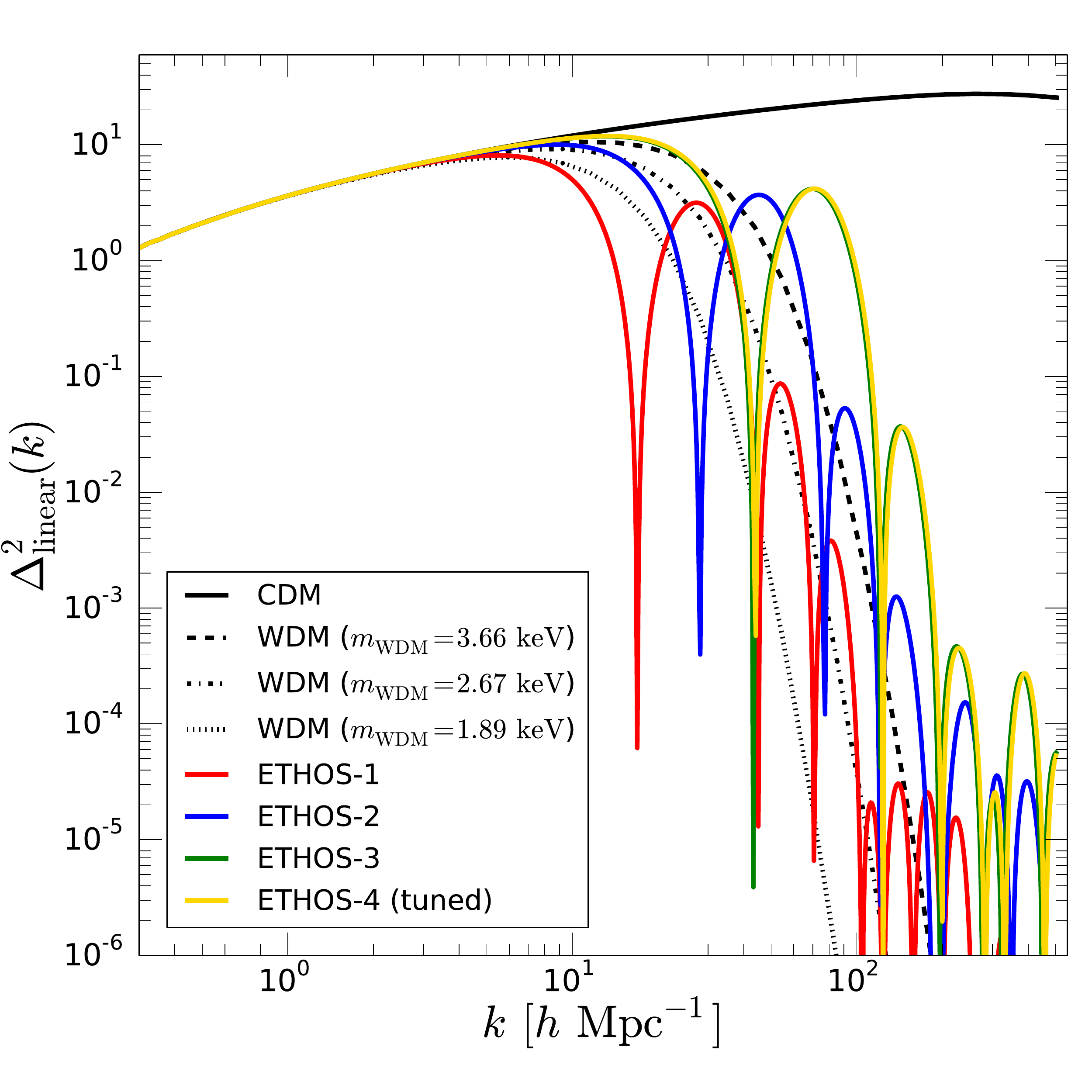}
\includegraphics[width=0.475\textwidth]{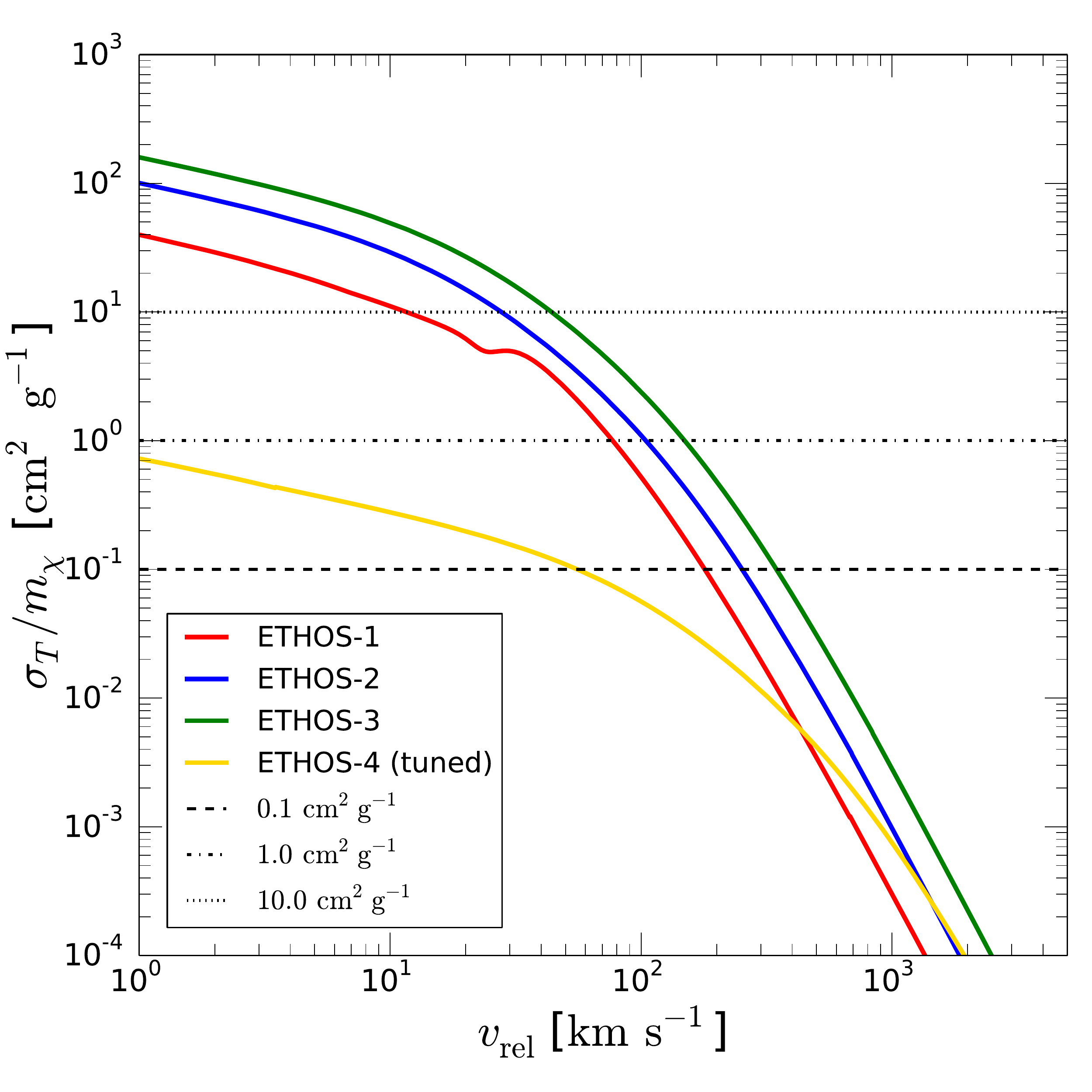}
\caption{Properties of the effective DM models relevant for structure
formation. Left: Linear initial matter power spectra ($\Delta_{\rm
linear}(k)^2=k^3 P_{\rm linear}(k)/2\pi^2$) for the different models (CDM and
ETHOS models ETHOS-1 to ETHOS-4) as a function of comoving wavenumber $k$.  The ETHOS
models differ in the strength of the damping and the dark acoustic
oscillations at small scales. As a reference, we also include thermal-relic-WDM
models, which are close to each model in ETHOS. Right: Velocity dependence of the transfer
cross-section per units mass ($\sigma_T/m$) for the different ETHOS models.
Models ETHOS-1 to ETHOS-3 have $\sigma_T/m\propto v_{\rm rel}^{-4}$ for large relative
velocities. For low velocities the cross sections can be as high as $100\,{\rm
cm}^2\,{\rm g}^{-1}$. }
\label{fig:models}
\end{figure*}

For a significant fraction of such DM models, the amplitude and shape of the
initial linear matter power spectrum, and the size of the self-interaction
cross section at different velocities largely determine the abundance and
structure of DM halos on a variety of mass scales. Therefore, distinct DM
particle models that make similar predictions for the linear matter power
spectrum and self-interaction cross section, can be classified into a single
``effective theory of structure formation'' (ETHOS). This is useful since all
DM particle models that map to a given effective theory can be simultaneously
constrained by comparing a single ETHOS numerical simulation to observations
at no extra cost or effort.  This ETHOS framework aims at generalising the
theory of DM structure formation to include a wide range of allowed DM
phenomenology \citep[see][for details]{Cyr-Racine2015}.  The goal of
the present work is to explore a few ETHOS benchmark cases and demonstrate their
interesting behaviour at the scale of galactic subhaloes. We will also
demonstrate that the parameter space of these models can actually be
constrained by astrophysical observations; for example, in cases that lead to
an unrealistic reduction of substructure in galactic haloes, or inner density
profiles that are too low compared to astrophysical constraints.

This paper has the following structure. We present the different ETHOS models
discussed in this work in Section 2, where we present their particle physics
parameters and their mapped initial linear power spectra and cross sections. We
introduce in total four different ETHOS models. Section 3 discusses the
different simulations carried out to explore these models. Here we describe our
uniform box and zoom-in simulations. The results of these simulations are then
presented in Section 4, where we first focus on general results based on the
uniform box simulations before discussing the structure of the galactic halo
within the different models. In this section we will also try to construct a
model which solves some of the outstanding small-scale problems of the MW
satellites. Finally, we present our summary and conclusions in Section 5.

\section{Effective Models}

\begin{table*}
\begin{center}
\begin{tabular}{llllllllllll}
\hline
\!\!\!\!Name \!\!\!\!\!\!\!  & $\alpha_\chi$ \!\!\!\!\!\!\!  & $\alpha_\nu$ \!\!\!\!\!\!\!  & $m_\phi$              \!\!\!\!\!\!\! & $m_\chi$               \!\!\!\!\!\!\!\!  &  $r_{\rm DAO}$       \!\!\!\!\!\!\!\!  & $r_{\rm SD}$ & $a_4$     &   $\langle\sigma_{T}\rangle_{\rm 30}/m_\chi$ & $\langle\sigma_{T}\rangle_{\rm 200}/m_\chi$ & $\langle\sigma_{T}\rangle_{\rm 1000}/m_\chi$  \\
\!\!\!\!                     &               \!\!\!\!\!\!\!  &              \!\!\!\!\!\!\!  & $[{\rm MeV\,c^{-2}}]$ \!\!\!\!\!\!\! & $[{\rm GeV\,c^{-2}}]$  \!\!\!\!\!\!\!\!  &  $[h^{-1}\,{\rm Mpc}]$ \!\!\!\!\!\!\!\!  & $[h^{-1}\,{\rm Mpc}]$ & [$h\,{\rm Mpc}^{-1}$] &  $[{\rm cm}^2\,{\rm g}^{-1}]$ & $[{\rm cm}^2\,{\rm g}^{-1}]$  & $[{\rm cm}^2\,{\rm g}^{-1}]$ & \\
\hline
\hline
\!\!\!\!CDM         \!\!\!\!\!\!\!   & --            \!\!\!\!\!\!\!  & --           \!\!\!\!\!\!\!  & --                    \!\!\!\!\!\!\! & --                     \!\!\!\!\!\!\!\!  & --                   \!\!\!\!\!\!\!\!  & --                          & -- & -- & -- & \\
\!\!\!\!ETHOS-1          \!\!\!\!\!\!\!   & $0.071$       \!\!\!\!\!\!\!  & $0.123$      \!\!\!\!\!\!\!  & $0.723$               \!\!\!\!\!\!\! & $2000$                 \!\!\!\!\!\!\!\!  & $0.362$              \!\!\!\!\!\!\!\!  & $0.225$                &  14095.65   & 4.98   & 0.072  & 0.0030 \\
\!\!\!\!ETHOS-2          \!\!\!\!\!\!\!   & $0.016$       \!\!\!\!\!\!\!  & $0.03$       \!\!\!\!\!\!\!  & $0.83$                \!\!\!\!\!\!\! & $500$                  \!\!\!\!\!\!\!\!  & $0.217$              \!\!\!\!\!\!\!\!  & $0.113$                &  1784.05  & 9.0    & 0.197 & 0.00097 \\
\!\!\!\!ETHOS-3          \!\!\!\!\!\!\!   & $0.006$       \!\!\!\!\!\!\!  & $0.018$      \!\!\!\!\!\!\!  & $1.15$                \!\!\!\!\!\!\! & $178$                  \!\!\!\!\!\!\!\!  & $0.141$              \!\!\!\!\!\!\!\!  & $0.063$                &  305.94  & 16.9   & 0.48 & 0.0028 \\
\!\!\!\!ETHOS-4 (tuned)  \!\!\!\!\!\!\!   & $0.5$         \!\!\!\!\!\!\!  & $1.5$        \!\!\!\!\!\!\!  & $5.0$                \!\!\!\!\!\!\! & $3700$                  \!\!\!\!\!\!\!\!  & $0.138$              \!\!\!\!\!\!\!\!  & $0.0615$               &  286.09  & 0.16     & 0.022 & 0.00075 \\
\hline
\end{tabular}
\begin{tabular}{llllll}
CDM: &  has MS and TBTF problems  & ETHOS-1: & over-solves MS and TBTF problems & ETHOS-2: & over-solves TBTF problem \\
 ETHOS-3: & over-solves TBTF problem & ETHOS-4 (tuned): & alleviates MS and TBTF problems\\
\end{tabular}
\end{center}
\caption{Parameters of the effective DM models considered in this paper. We
study in total five different scenarios (CDM and models ETHOS-1 to
ETHOS-4).   ETHOS models are characterised by four intrinsic parameters: the coupling
between the mediator and DM ($\alpha_\chi\equiv g_\chi^2/4\pi$), the coupling between the mediator
and a massless neutrinos-like fermion ($\alpha_\nu\equiv g_\nu^2/4\pi$), the mediator mass ($m_\phi$), and the mass of
the DM particle ($m_\chi$). In principle, the ratio of neutrino-like fermion and photon
temperature constitutes another parameter that follows from the underlying
particle physics framework; for definiteness, we will set it throughout to
$0.5$ in this paper.  ETHOS models are characterised by five effective parameters (see Eqs.~\ref{eff_par}$-$\ref{eff_par_red}): $a_4$ and $\alpha_{l\geq2}=3/2$ (constant for these models and thus absent from the table), which determine 
the linear power spectrum (left panel of Fig.~\ref{fig:models}), while the velocity dependence of the DM self-interaction cross section (right panel of Fig.~\ref{fig:models})
is described by three characteristic cross sections at three different velocities ($30,200,1000\kms$).
As a reference, we also provide two characteristic comoving length
scales: the DM sound horizon ($r_{\rm DAO}$), and the Silk damping scale
($r_{\rm SD}$). 
ETHOS-4 is a tuned model that was
specifically set up to address some of the small-scale issues of CDM (the MS problem
and the TBTF problem).}
\label{table:models}
\end{table*}

The different DM models that we investigate in this paper are summarised in
Table~\ref{table:models}.  For all simulations we use the following cosmological
parameters:  $\Omega_m=0.302$, $\Omega_{\Lambda}=0.698$, $\Omega_b=0.046$,
$h=0.69$, $\sigma_8=0.839$ and $n_{s}=0.967$, which are consistent with recent
Planck data \citep[][]{Planck2014, Spergel2015}. We study mainly five different
DM models, which we label CDM and ETHOS-1 to ETHOS-4. 
In the parameter space of ETHOS, these
models are represented by a specific transfer function (see left panel of
Fig.~\ref{fig:models} for the resulting linear dimensionless power spectra),
and a specific velocity-dependent transfer cross-section for DM (see right
panel of Fig.~\ref{fig:models}). Our discussion will mostly focus on ETHOS-1 to ETHOS-3,
which demonstrate the basic features of our ETHOS models. ETHOS-4 is a tuned model
that was specifically set up to address the small-scale issues of CDM (the
MS problem and the TBTF problem). We discuss this model
towards the end of the paper.

These models arise within the effective framework of ETHOS, described in detail in
\cite{Cyr-Racine2015}, which we summarise in the following. ETHOS provides a mapping between the intrinsic parameters (couplings, masses, etc.)
defining a given DM particle physics model, and (i) the effective parameters controlling the shape of the linear matter power spectrum, and (ii) the 
effective DM transfer cross section ($\langle\sigma_T\rangle/m_{\chi}$); both at the relevant scales for structure formation. Schematically:
\begin{eqnarray}\label{eff_par}
\Big\{m_\chi, \{g_i\},\{h_i \},\xi \Big\} &\!\!\!\!\!\!\!\rightarrow\!\!\!\!\!\!\!& \Big\{\omega_{\rm DR},\{a_n,\alpha_l\},\{b_n,\beta_l\}, \{d_n, m_\chi,\xi\}\Big\}\nonumber\\&\rightarrow& P_{\rm lin, matter}(k)\nonumber\\
\Big\{m_\chi,\{h_i\}, \{g_i\} \Big\} &\!\!\!\!\!\!\!\rightarrow\!\!\!\!\!\!\!& \Bigg\{\frac{\langle \sigma_T \rangle_{30}}{m_\chi},\frac{\langle \sigma_T \rangle_{220}}{m_\chi},\frac{\langle \sigma_T \rangle_{1000}}{m_\chi}\Bigg\},
\end{eqnarray}
where the parameters on the left are the intrinsic parameters of the DM model: $m_\chi$ is the mass of the DM particle, $\{g_i\}$ represents the set of coupling constants, $\{h_i\}$ is a set of other internal parameters such as mediator mass and number of degrees of freedom, and $\xi = (T_{\rm DR}/T_{\rm CMB})|_{z=0}$ is the present day DR to CMB temperature ratio. 

The effective parameters of the framework are on the right of Eq.~\ref{eff_par}, which in all generality include the cosmological density of DR $\omega_{\rm DR} \equiv \Omega_{\rm DR} h^2$, the set $\{a_n,\alpha_l\}$ characterising the DM-DR interaction, the $\{b_n,\beta_l\}$ set characterising the presence of DR self-interaction (relevant, for instance, to non-abelian DR), and the parameter set $\{d_n, m_\chi,\xi\}$ determining the evolution of the DM temperature and adiabatic sound speed. This latter quantity is very small for non-relativistic DM, and it has thus little impact on the evolution of linear DM perturbations (except on very small scales, irrelevant for galaxy formation/evolution). In this work, we focus our attention on the effect of DM-DR interaction on the evolution of DM perturbations. The physics of these effects are captured by the parameters $\{a_n,\alpha_l\}$, where the set of $l-$dependent coefficients $\alpha_l$ encompasses information about the angular dependence of the DM-DR scattering cross section, whereas the $a_n$ are the coefficients of the power-law expansion in temperature (redshift) of the DM drag opacity caused by the DM-DR interaction \citep[see Section II E of][]{Cyr-Racine2015}. Physically, a single non-vanishing $a_n$ implies that the squared matrix element for the DM-DR scattering process scales as $|\mathcal{M}|^2\propto (p_{\rm DR}/m_\chi)^{n-2}$, where $p_{\rm DR}$ is the DR momentum. We leave the impact of DR self-interactions on the matter power spectrum to a future study. We note that DR self-interaction as parametrised by $\{b_n,\beta_l\}$ can actually have a non-negligible effect on the linear matter power spectrum through its influence on the gravitational shear stress. However, this latter effect is generally subdominant compared to the DM-DR interactions studied in the present work. 

The other set of effective parameters in ETHOS are related to DM self-scattering. Although each particle physics model would have a specific transfer cross section, in ETHOS we classify (characterise) a given model based on the values of its cross section at three relative velocities, those characteristic of dwarf galaxies ($\sim30$~km s$^{-1}$), the Milky-Way-size galaxies ($\sim220$~km s$^{-1}$) and galaxy clusters ($\sim1000$~km s$^{-1}$).\footnote{Note that in some cases one needs to go beyond the transfer cross section to describe the effect of self-interactions, see~e.g.~\cite{Kahlhoefer:2013dca}.}
 The choice of these three characteristic velocities is arbitrary but it allows us at a glance to (i) check whether a given model is compatible with observations, and (ii) have a reliable estimate at what the outcome of the simulation of a given model would be based on the results of models already simulated, which have similar values of the transfer cross section. For instance, if two models have the same values of $\langle\sigma_T \rangle_{30}/{m_\chi}$, full simulations of isolated dwarfs in each model are likely to yield similar results, even though they might have very different values of $\langle\sigma_T \rangle_{1000}/{m_\chi}$. Furthermore, these characteristic velocities mark also three relevant regimes for any model containing DM self-interactions: (i) the dwarf-scale regime where the CDM model is being challenged, and where the transfer cross section is largely unconstrained, (ii) the intermediate-scale regime where a large cross section can lead to the evaporation of subhaloes in Milky-Way-size galaxies, and (iii) the cluster-scale regime where observations put the strongest constraints to the cross section.

The ETHOS framework described above is general, but for the purpose of this work we restrict ourselves to an underlying particle physics model which assumes, like in \cite{vandenAarssen2012}, a
massive fermionic DM particle ($\chi$) interacting with a massless neutrino-like fermion ($\nu$)
via a massive vector mediator ($\phi$). This model is characterised by an interaction between DM and
DR and DM-DM self-interactions \citep[see Section II F.1 of][for details]{Cyr-Racine2015}. The former gives rise to the features in the power spectrum, which are absent in ordinary CDM transfer functions, while the latter alters the evolution of dark matter haloes across time. This model is characterised by a squared matrix element scaling as $(p_{\rm DR}/m_\chi)^2$, which immediately implies that the impact of DM-DR scattering on the linear matter power spectrum is entirely captured by a non-vanishing $a_4$ coefficient. 
For DM-DR interactions leading to late kinetic decoupling, this is indeed a very commonly encountered 
situation according to a recent comprehensive classification of such scenarios
\cite{Bringmann:2016ilk}; note, however, that in the presence of \emph{scalar} mediators it is sometimes 
rather $a_2$ that is the only non-vanishing coefficient $a_n$ (depending on the spin of DM and DR).

In our case, the ETHOS mapping is reduced to: 
\begin{eqnarray}\label{eff_par_red}
\Big\{m_\chi, m_\phi,g_\chi,g_\nu,\eta_\chi,\eta_\nu,\xi\Big\}\longrightarrow\hspace{3.8cm}\\
\Bigg\{\omega_{\rm DR},a_4,\alpha_{l\geq2}=\frac{3}{2},\frac{\langle \sigma_T \rangle_{30}}{m_\chi},\frac{\langle \sigma_T \rangle_{220}}{m_\chi},\frac{\langle \sigma_T \rangle_{1000}}{m_\chi}\Bigg\}\nonumber.
\end{eqnarray}
The model is characterised by  
six intrinsic particle physics parameters: the mass of the DM particle
($m_\chi$), the mediator mass ($m_\phi$), the coupling between the mediator and DM
($g_\chi$), the coupling between the mediator and neutrino-like fermions ($g_\nu$), the number of DM spin states $(\eta_\chi)$, and the number of spin states of the neutrino-like fermion $(\eta_\nu)$.  In principle, the ratio of neutrino-like fermion and photon temperature $\xi$ constitutes another parameter that follows from the underlying particle physics framework; for definiteness, we will set it throughout to $0.5$ in this work. The effective ETHOS parameters that fully characterise the linear power spectrum are then reduced to three: the abundance of DR $\omega_{\rm DR}$, the opacity parameter $a_4$ ($a_{n\neq4}=0$), and a set of constant $\alpha_{l\geq2}$ values. It is possible to calculate these parameters analytically \citep[][]{Cyr-Racine2015}
\begin{eqnarray}
a_4&=&(1+z_{\rm D})^{4}\frac{\pi g_\chi^2g_\nu^2}{m_\phi^4} \frac{ \tilde{\rho}_{\rm crit} }{ m_\chi} \left(\frac{310}{441}\right) \xi^2 T_{\rm CMB,0}^2,\nonumber\\
\alpha_{l\geq2}&=&\frac{3}{2},
\end{eqnarray}
where $\tilde{\rho}_{\rm crit}  \equiv  \rho_{\rm crit}/h^2$ with $\rho_{\rm crit}$ the critical density of the Universe, and $T_{\rm CMB, 0}$ is the temperature of the CMB today. The normalization redshift $z_{\rm D}$ is arbitrary, but choosing it to be the redshift of DM kinetic decoupling ensures that the $a_n$ coefficients are generally of order unity. For models that modify the linear matter power spectrum on sub-galactic scales, we usually have $z_{\rm D}\gtrsim 10^7$. The generic form of the $a_4$ coefficient is easy to understand: the combination $g_\chi^2g_\nu^2/m_\phi^4$ is the leading factor in the squared scattering amplitude, the term $\tilde{\rho}_{\rm crit}/m_\chi$ is proportional to the DM number density, and the remaining factors come from thermally averaging $p_{\rm DR}^2$ over the Fermi-Dirac distribution describing the DR phase-space. The non-unity value of the angular coefficients $\alpha_{l\geq2}$ is caused by the angular dependence of the DM-DR scattering cross section which scales as $1+\cos{\theta}$, where $\theta$ is the angle between the outgoing DM and DR particles.

The other set of effective parameters are those related to the velocity-dependent DM self-interaction cross section expected for a Yukawa potential between the DM particles (given the choice of particle physics model; see e.g. \citealt{Loeb2011}), evaluated at the three characteristic velocities of the ETHOS framework. The formula for the DM momentum transfer cross section has been obtained in analogy with the screened Coulomb scattering in a plasma \citep[see the improved Eqs. 100-101 in][]{Cyr-Racine2015}.  Most
importantly, the behaviour of the cross sections is to fall off rapidly towards
large relative velocities. It is therefore possible to have large
cross-sections at small scales (i.e., low relative velocities, for example at
the dwarf scale) while at the same time satisfy constraints on cluster scales.

Table~\ref{table:models} specifies the relevant particle physics and effective parameters for the cases we have simulated in this work. The effective parameters that control the shape of the linear power spectrum are
related to more familiar scales in the initial power spectrum: the comoving  diffusion (Silk) damping scale ($r_{\rm SD}$) and the
DM comoving sound horizon ($r_{\rm DAO}$). These are generic scales, which occur
in models where DM is coupled to relativistic particles in the early
Universe, i.e., they are not only a consequence of the specific particle physics scenario used here.
Currently, or simulations only cover the regime
for which $r_\mathrm{DAO} \gtrsim r_\mathrm{SD}$ (``weak'' DAOs); for an example of a simulation in the strong
DAOs regime, with $r_\mathrm{DAO} \gg r_\mathrm{SD}$, see~\cite{Buckley2014}.

As a reference, the left panel of Fig.~\ref{fig:models} also shows three WDM
power spectra for thermal relics, which are described by a sharp cut-off
\citep[we follow][with $\nu=1$]{Bode2001}:
\begin{equation}
P_{\rm WDM}(k) = T^2(k) P_{\rm CDM}(k), \quad T(k) = (1 + (\alpha k)^{2})^{-5},
\end{equation}
where the $\alpha$ parameter defines the cutoff scale in the
initial power spectrum and is related to the free-streaming of WDM particles.
The $\alpha$ value can be associated with a generic thermal relic WDM particle
mass, $m_{\rm WDM}$, using the relation \citep[][]{Bode2001}:
\begin{equation}
\alpha \!\! = \!\! \frac{0.05}{h\,{\rm Mpc}^{-1}} \!\! \left(\frac{m_{\rm WDM}}{1\,{\rm keV}}\right)^{\!\!-1.15} \!\!\! \left(\frac{\Omega_{\rm WDM}}{0.4}\right)^{\!\!\!0.15} \!\!\!\! \left(\frac{h}{0.65}\right)^{\!\!\!1.3} \!\!\!\! \left(\frac{g_{\rm WDM}}{1.5}\right)^{\!\!-0.29},
\end{equation}
where $\Omega_{\rm WDM}$ is the WDM contribution to the density parameter, and
$g_{\rm WDM}$ the number of degrees of freedom. It is conventional to use $1.5$
as the fiducial value for $g_{\rm WDM}$ for the WDM particle. The left panel of
Fig.~\ref{fig:models} shows also the WDM particle masses for the three cases,
which were chosen by eye to match the initial power decline of the ETHOS
models as well as the FoF halo mass function (see Fig.~\ref{fig:massfct} and discussion further down).

We note that the Lyman-$\alpha$ forest is sensitive to any sort of small-scale
cutoffs in the power spectrum; a feature that puts, for example, tight
constraints on the mass of thermal-relic-WDM particles \citep{Viel2013}.  The
acoustic oscillation ($r_{\rm DAO}$) and damping ($r_{\rm SD}$) scales can
therefore, in principle, be constrained via Lyman-$\alpha$ forest data as well.
Since the shape of the cutoff in our  models is very different from the
exponential cutoff in WDM models, it is thus necessary to perform detailed
hydrodynamical simulations for the models presented here in order to obtain appropriate
Lyman-$\alpha$ forest constraints. We will discuss this in a forthcoming work
(Zavala et al., in prep).

\section{Simulations}

We generate initial conditions at $z=127$ within a $100\hmpc$ periodic box (our
parent simulation) from which we select a MW-size halo to be resimulated with a
zoom technique. The transfer functions for all DM models were generated with a
modified version of the CAMB code~\citep[][]{Seljak1996,Lewis2011}, as
described in \cite{Cyr-Racine2015}.  All initial conditions were
generated with the MUSIC code~\citep[][]{Hahn2011}.  The uniform parent
simulation is performed at a resolution of $1024^3$ particles yielding a DM particle mass
resolution of $7.8 \times 10^7\hmsun$ and a spatial resolution
(Plummer-equivalent softening length) of $\epsilon=2\hkpc$. This is sufficient
to resolve haloes down to $\sim 2.5 \times 10^9\hmsun$ with about $32$
particles.  We note that the mass and spatial resolution of this  parent
simulation is slightly better than the simulations presented in
\cite{Buckley2014}, which have a smaller simulation volume. The parent
simulation presented here has therefore better statistics and also includes
more massive clusters. It contains $10$ haloes with a virial mass ($M_{\rm
200,crit}$) above $10^{14}\hmsun$ at $z=0$.  

The galactic halo for resimulation was randomly selected from a sample of
haloes that have masses between $1.58 \times 10^{12}\msun$ and 
$1.61\times10^{12}\msun$, which is in the upper range of current estimates for the mass of
the MW halo \citep[see Fig. 1 of][]{Wang2015}. This sample was created using
only those MW-size haloes which do not have another halo more massive than half
their masses within $2\hmpc$ (this is a criterion for isolation). We stress
that we do not consider a local group analog here in this first study. We have
simulated the selected halo at three different resolutions, level-3 to level-1,
which are summarised in Table~\ref{table:resolution}.  For these resimulations,
the softening length is fixed in comoving coordinates until $z=9$, and is then
fixed in physical units until $z=0$.  The latter value is quoted in
Table~\ref{table:resolution}. The number of high resolution particles refers to
the CDM simulation only; the other DM models produce slightly different numbers. 
The most basic characteristics of the halo
are presented in Table~\ref{table:halo} for the highest resolution simulations. 

Self-scattering of DM particles was implemented into the {\tt AREPO}
code~\citep[][]{Springel2010} following the probabilistic approach described
in~\cite{Vogelsberger2012}, which assumes that scattering is elastic and
isotropic. This implementation has previously been used, in the context of
standard SIDM (i.e. with the same power spectrum as CDM), to constrain the
self-interaction cross section at the scale of the MW dwarf
spheroidals~\citep[][]{Zavala2013}, predict direct detection signatures of
self-interactions~\citep[][]{Vogelsberger2013}, and study the impact on lensing signals~\citep[][]{Vegetti2014}. It was also used to find that
self-interactions can leave imprints in the stellar distribution of dwarf
galaxies by performing the first SIDM simulation with baryons presented in
~\cite{Vogelsberger2014b}. 

\begin{table}
\begin{center}
\begin{tabular}{cccccc}
\hline
Name     &  $ m_{\rm DM}$ [${\rm M_\odot}$]         & $\epsilon$ [${\rm pc}$]    & $N_{\rm hr}$  \\
\hline     
\hline 
level-1  &  $ 2.756\times 10^4$                     & $72.4$                      & $444,676,320$   \\
level-2  &  $ 2.205\times 10^5$                     & $144.8$                     & $55,451,880$    \\
level-3  &  $ 1.764\times 10^6$                     & $289.6$                     & $7,041,720$     \\

\hline
\end{tabular}
\end{center}
\caption{Simulation parameters of the selected MW-size halo. We list the DM particle
mass ($m_{\rm DM}$), the Plummer-equivalent softening length ($\epsilon$), and the
number of high resolution particles ($N_{\rm hr}$). The softening length is
kept fixed in physical units for $z<9$. The number of high resolution
particles refers to the CDM case and slightly varies for the other DM models.}
\label{table:resolution}
\end{table}

\begin{table}
\begin{center}
\begin{tabular}{cccccccc}
\hline
Name        & $M_{\rm 200,crit}$          & $R_{\rm 200,crit}$    & $V_{\rm max}$                   & $R_{\rm max}$  & $N_{\rm sub}$  \\
            & $[10^{10}\,{\rm M_\odot}]$  & $[{\rm kpc}]$         & $[{\rm km}\,{\rm s}^{-1}]$     & $[{\rm kpc}]$  &                \\
\hline     
\hline 
CDM         & 161.28                     & 244.05                & 176.82                          & 68.29          & 16108          \\
ETHOS-1     & 160.47                     & 243.64                & 178.12                          & 62.58          & 590            \\
ETHOS-2     & 164.70                     & 245.75                & 181.49                          & 63.72          & 971            \\
ETHOS-3     & 163.36                     & 245.09                & 180.60                          & 64.37          & 1080           \\
ETHOS-4     & 163.76                     & 245.30                & 178.78                          & 69.18          & 1366           \\
\hline
\end{tabular}
\end{center}
\caption{Basic characteristics of the MW-size halo formed in the different DM models. We
list the mass ($M_{\rm 200,crit}$), radius ($R_{\rm 200,crit}$), maximum
circular velocity ($V_{\rm max}$), radius where the maximum circular velocity
is reached ($R_{\rm max}$), and the number of resolved subhaloes within $300\kpc$ ($N_{\rm
sub}$).}
\label{table:halo}
\end{table}

\section{Results}

\begin{figure}
\centering
\includegraphics[width=0.475\textwidth]{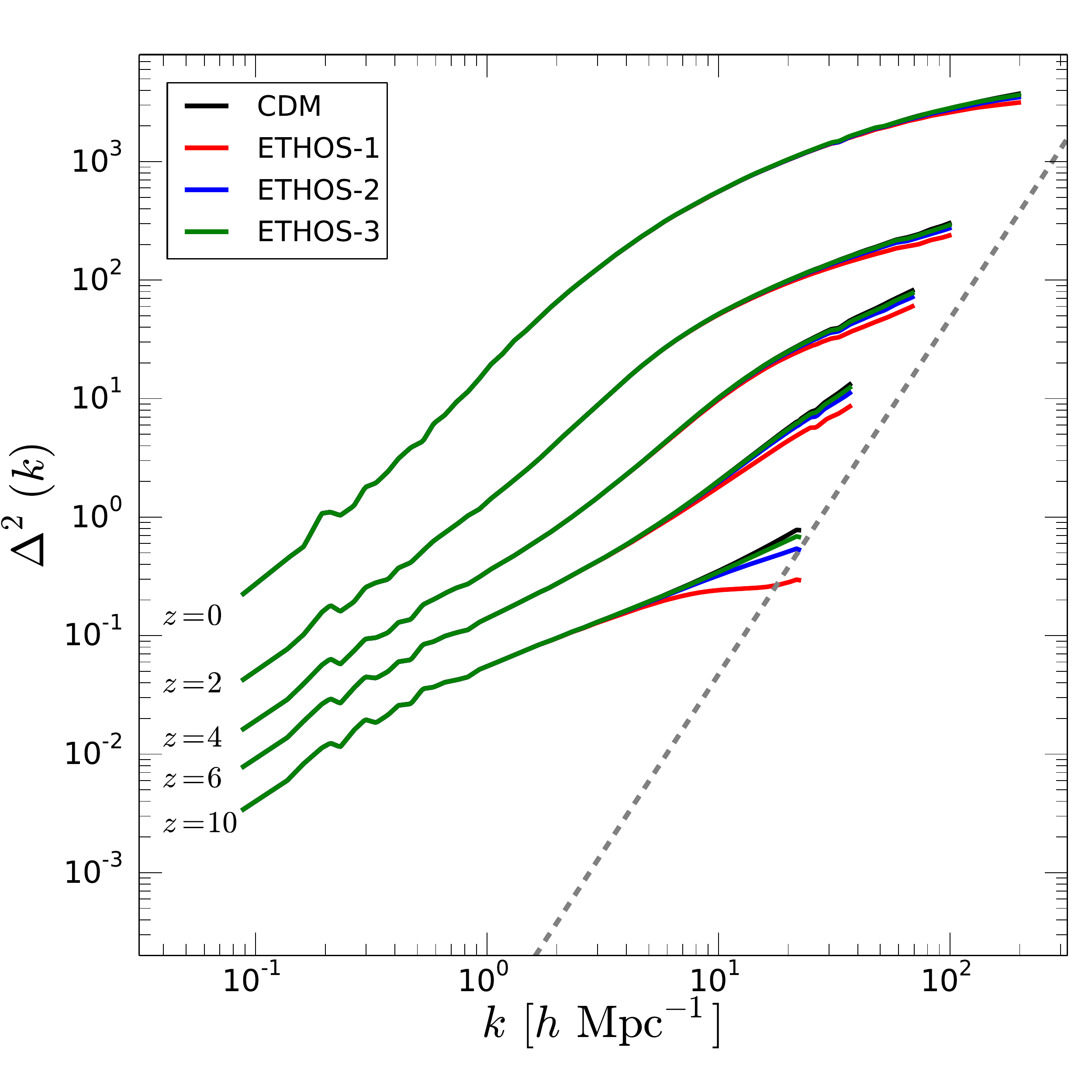}
\caption{Non-linear dimensionless power spectra, $\Delta(k)^2=k^3
P(k)/(2\pi^2)$, of the parent simulations for the different DM models at the
indicated redshifts ($z=10,6,4,2,0$). The dashed gray line denotes the
shot-noise limit expected if the simulation particles are a Poisson sampling
from a smooth underlying density field. The sampling is significantly
sub-Poisson at high redshifts and in low density regions, but approaches the
Poisson limit in nonlinear structures. The non-CDM models deviate significantly
from CDM at high redshifts, but this difference essentially vanishes towards
$z=0$.}
\label{fig:power}
\end{figure}

In the following, we first discuss some features of the large-scale
($100\hmpc$) parent simulations, followed by the main focus of our work, the
resimulated galactic halo. We show here only the results for CDM, and ETHOS-1 to ETHOS-3
since ETHOS-4 has the same initial power spectrum as ETHOS-3 and a significantly smaller
self-interaction cross section. The impact of SIDM effects on large scales is
thus much smaller for ETHOS-4 compared to ETHOS-1 to ETHOS-3. We have therefore not
performed a uniform box simulation for ETHOS-4.

\subsection{Large scale structure}

\begin{figure}
\centering
\includegraphics[width=0.475\textwidth]{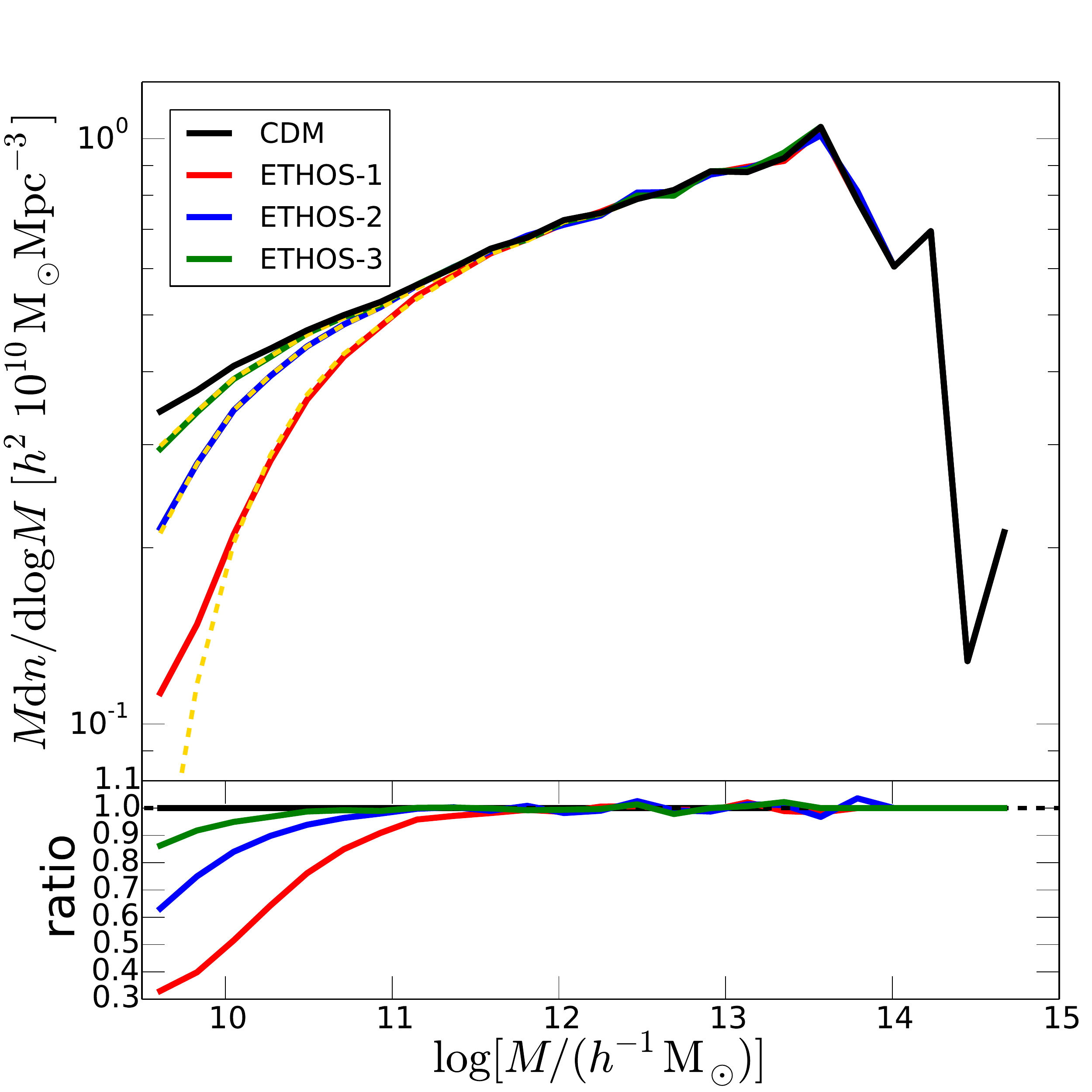}
\caption{Differential FoF halo mass function (multiplied by FoF mass squared) for the different DM models at $z=0$. Approximating the first DAO feature in the
linear power spectrum with a sharp power-law cutoff, we show the resulting
analytic estimates for the differential halo mass function of the different DM
models (yellow dashed). The lower panel shows the ratios between the different
simulation models relative to CDM.}
\label{fig:massfct}
\end{figure}

We first quantify the large scale distribution of matter in
Fig.~\ref{fig:power}, where we present the dimensionless power spectra,
$\Delta(k)^2=k^3 P(k)/(2\pi^2)$, at redshifts $z=10,6,4,2,0$ for our parent
simulations. The dashed gray line shows the shot noise power spectrum
caused by the finite particle number of the simulation, it gives an indication of the
resolution limit in this plot at low redshifts. The DAO features of the ETHOS-1 to
ETHOS-3 models, clearly visible on the primordial power spectrum (see left panel of
Fig.~\ref{fig:models}), are only preserved down to $z\sim10$ (where the first
oscillation is marginally resolved for model ETHOS-1). At lower redshifts, the
imprint of these features is significantly reduced and is essentially erased at
$z=0$.  At this time, although the power spectra of the non-CDM simulations are
relatively close to the CDM case, there is a slight suppression of power in the
ETHOS-1 to ETHOS-3 models for scales smaller than $k\gtrsim10^2~h {\rm Mpc}^{-1}$. This
suppression is largest for ETHOS-1 and smallest for ETHOS-3, which reflects the fact that
the initial power spectrum damping is largest for ETHOS-1 and smallest for ETHOS-3. Our
results therefore confirm the previous finding of~\cite{Buckley2014}, namely
that in the weak DAO regime, the non-linear evolution makes the differences
with CDM in the power spectra relatively small at low redshifts. We note that
we do not present images of the large scale density field since the different
models are indistinguishable on these scales.

\begin{figure}
\centering
\includegraphics[width=0.475\textwidth]{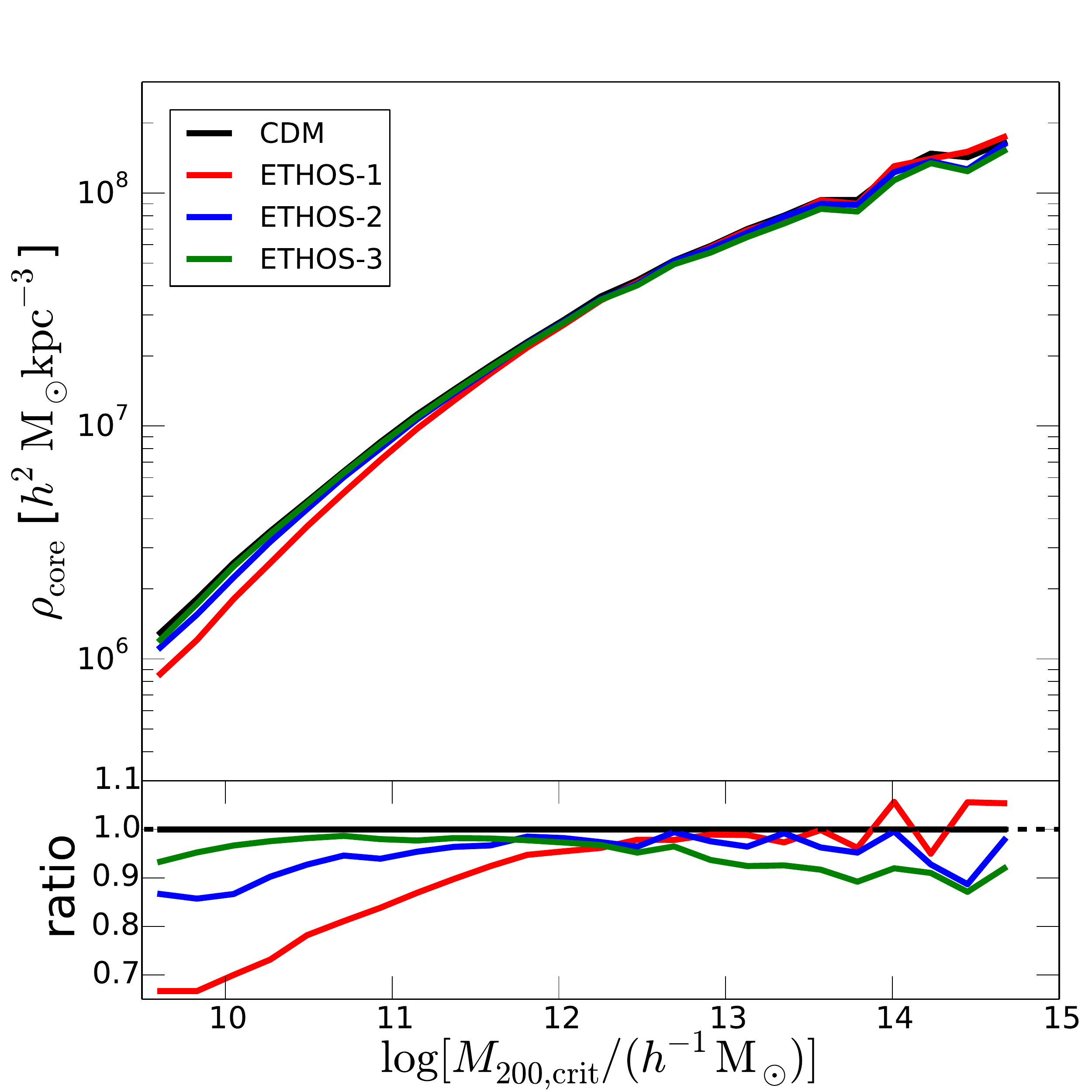}
\caption{Central density as a function of halo mass ($M_{\rm 200,crit}$) for
all main haloes (i.e. we do not include subhaloes here) for the different DM
models. The central (core) density is defined at $8.7\kpc$ (three times the
gravitational softening length).  The lower panel shows the ratio with respect
to the CDM case. ETHOS-1 and ETHOS-2 show decreasing core densities towards smaller halo
masses. Interestingly, ETHOS-3 shows a slightly different trend where the core
density compared to CDM is most reduced for the most massive haloes in the
simulation.}
\label{fig:cores}
\end{figure}

\begin{figure*}
\centering
\includegraphics[width=0.32\textwidth]{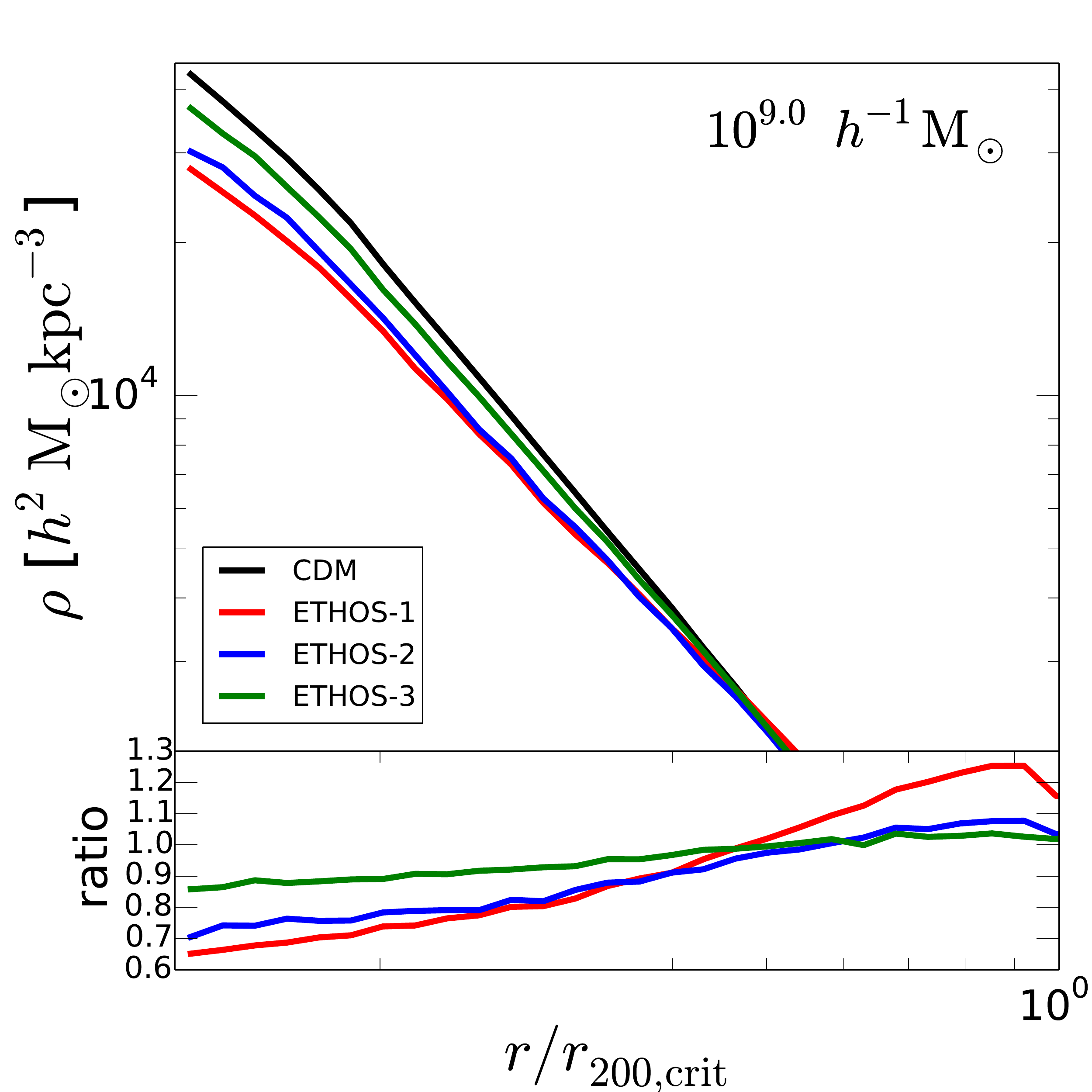}
\includegraphics[width=0.32\textwidth]{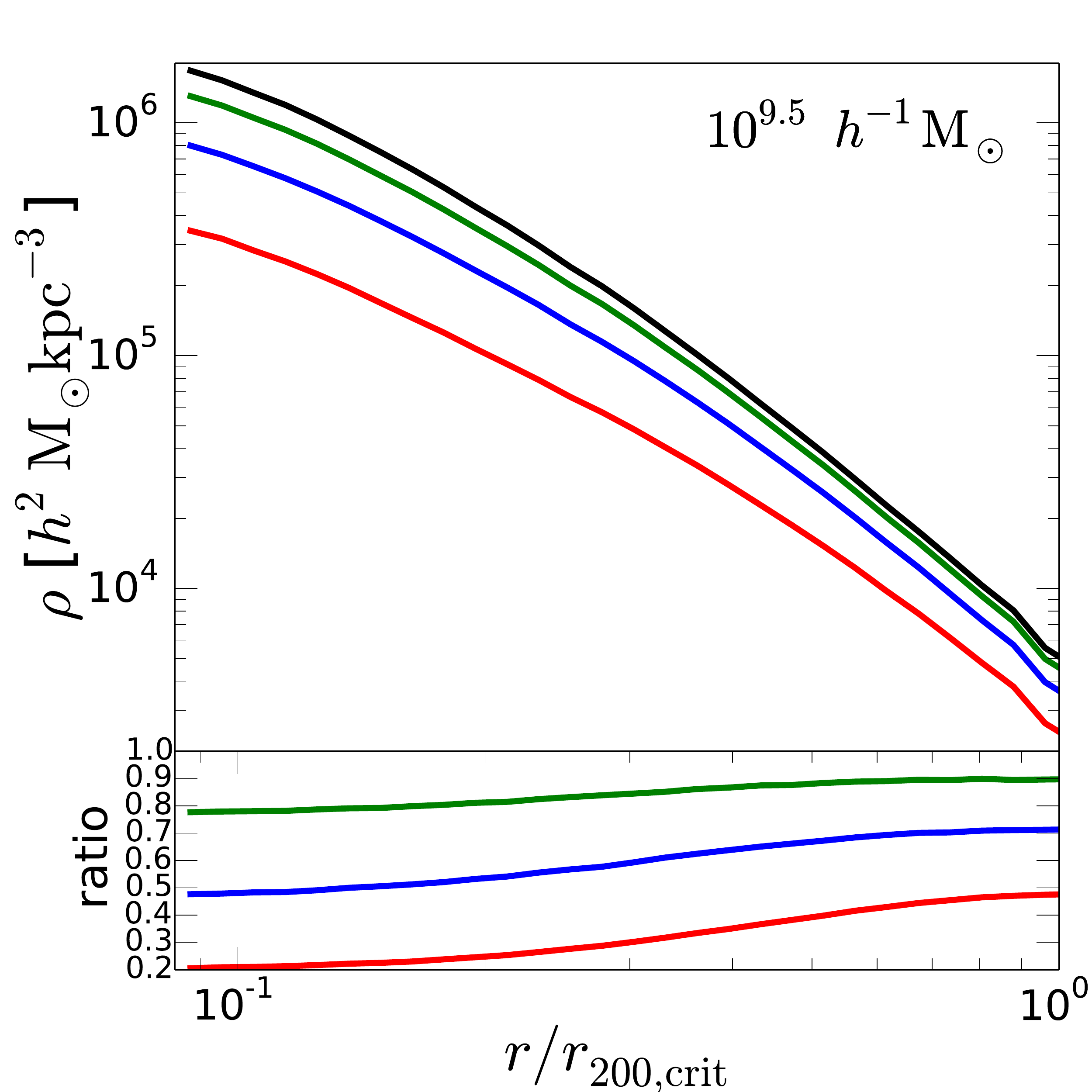}
\includegraphics[width=0.32\textwidth]{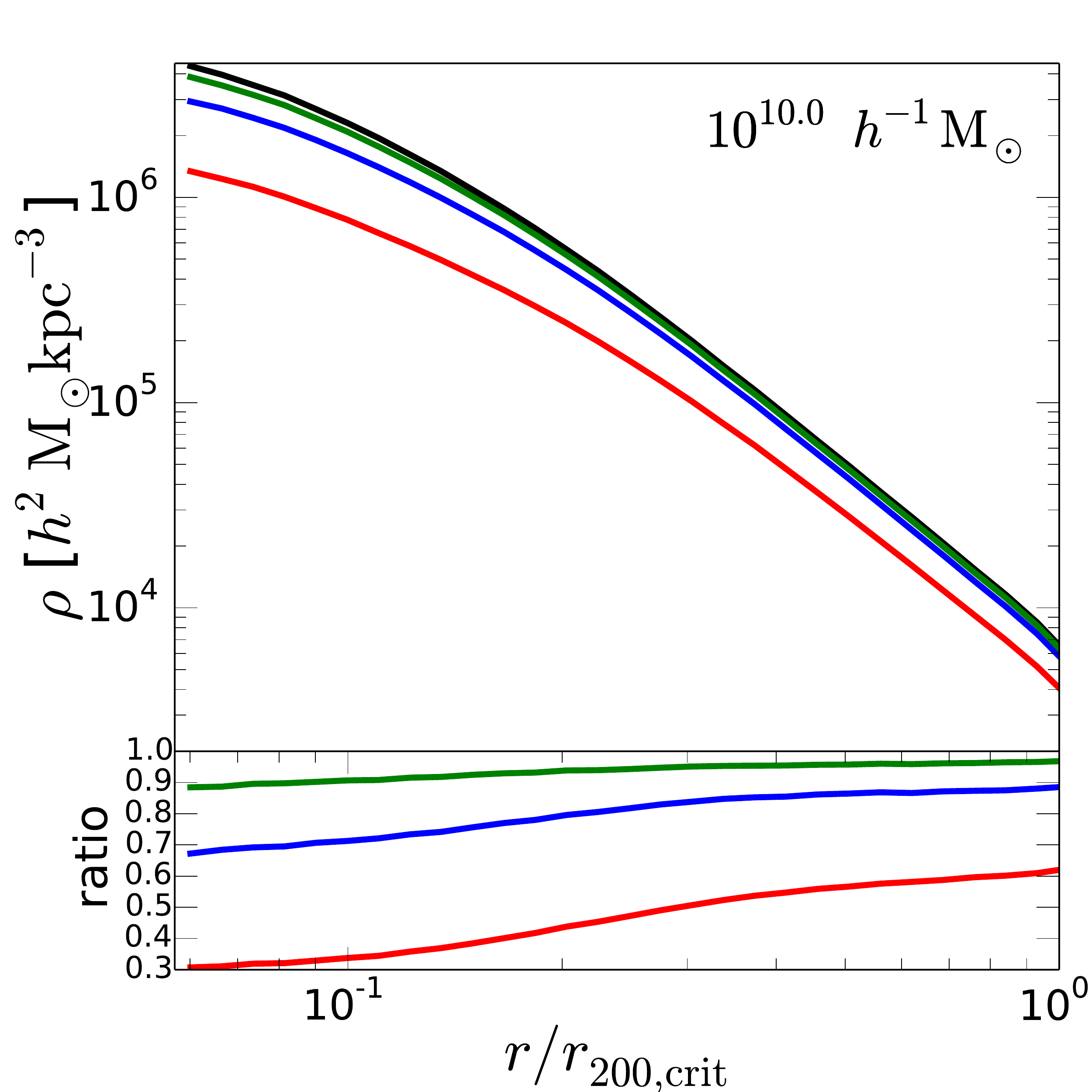}\\
\includegraphics[width=0.32\textwidth]{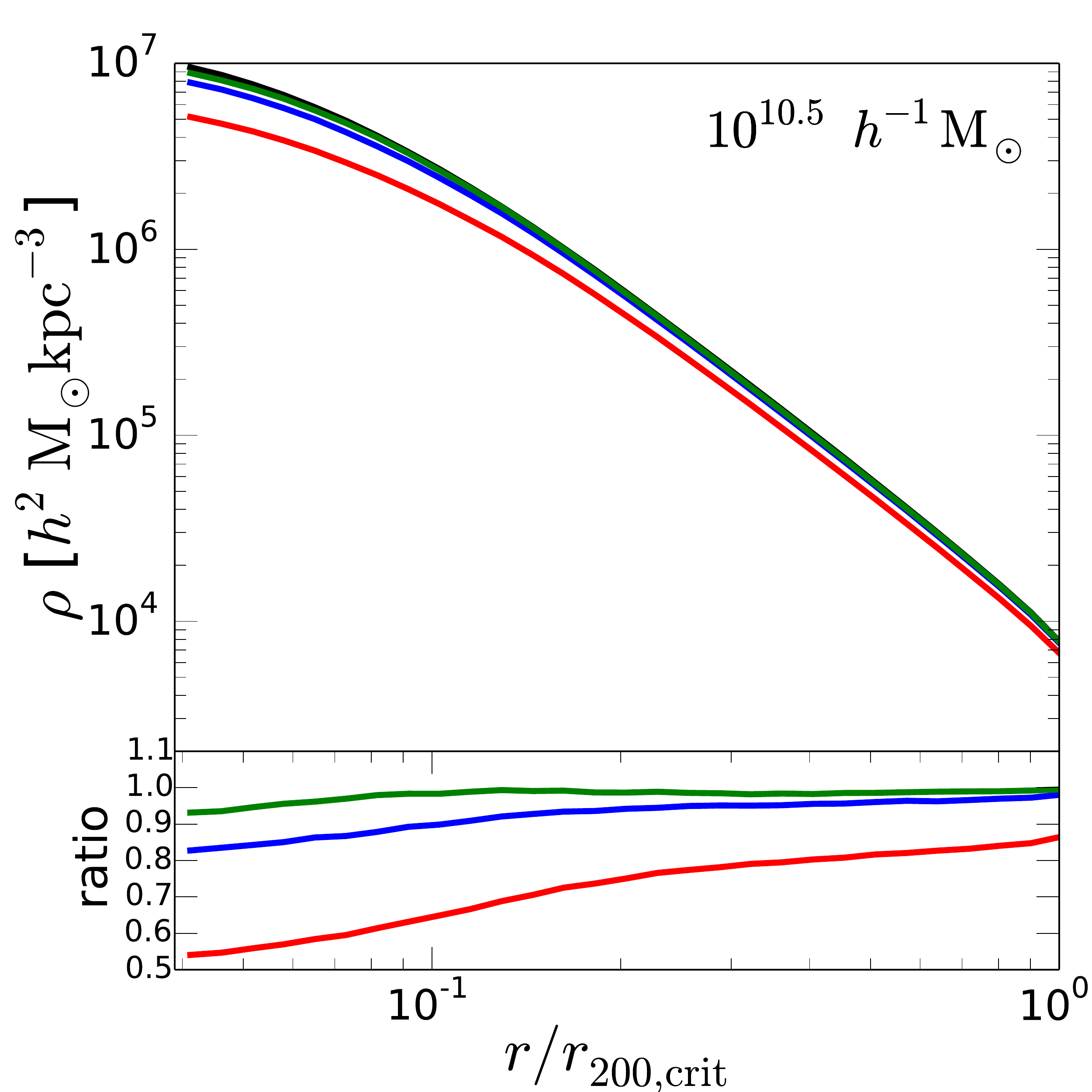}
\includegraphics[width=0.32\textwidth]{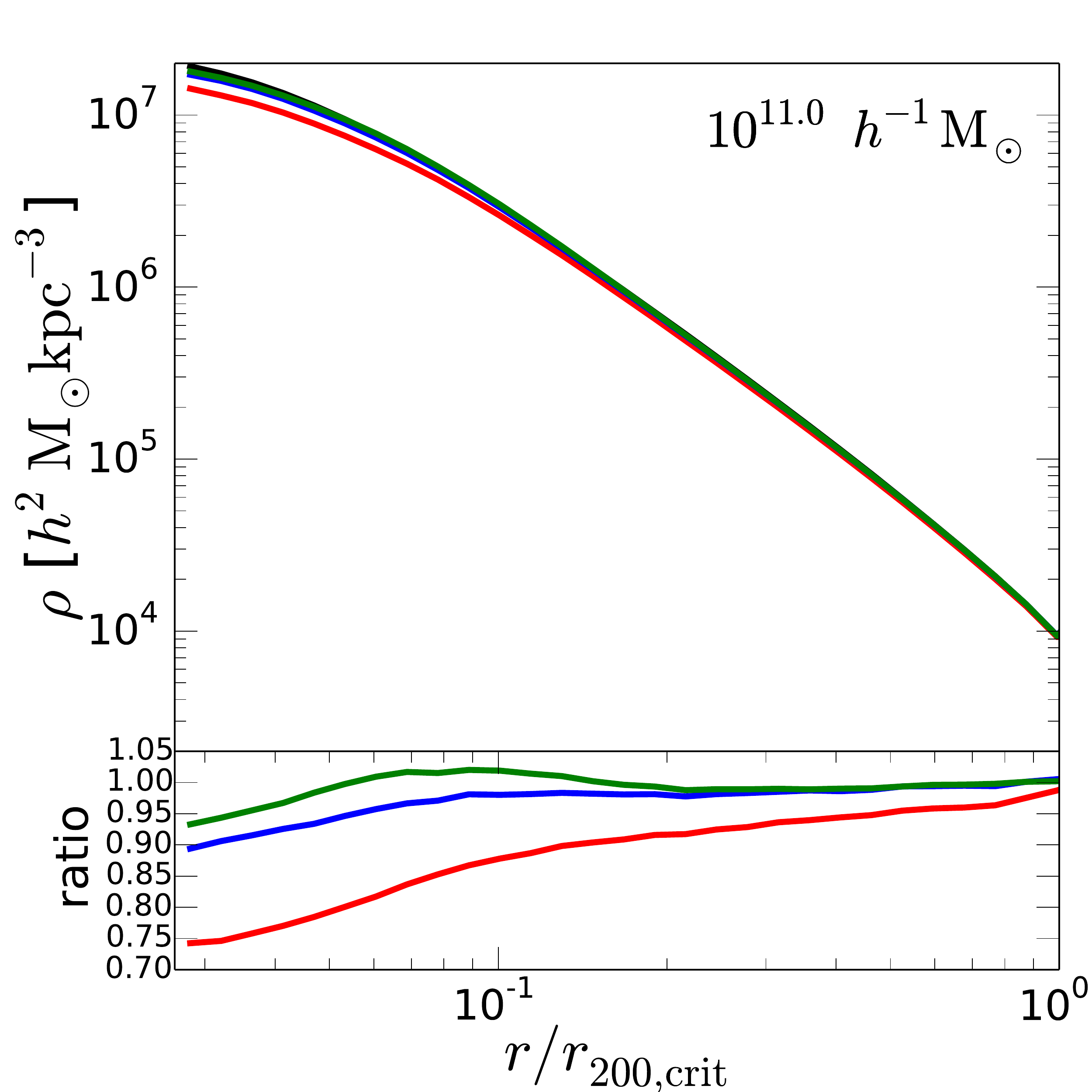}
\includegraphics[width=0.32\textwidth]{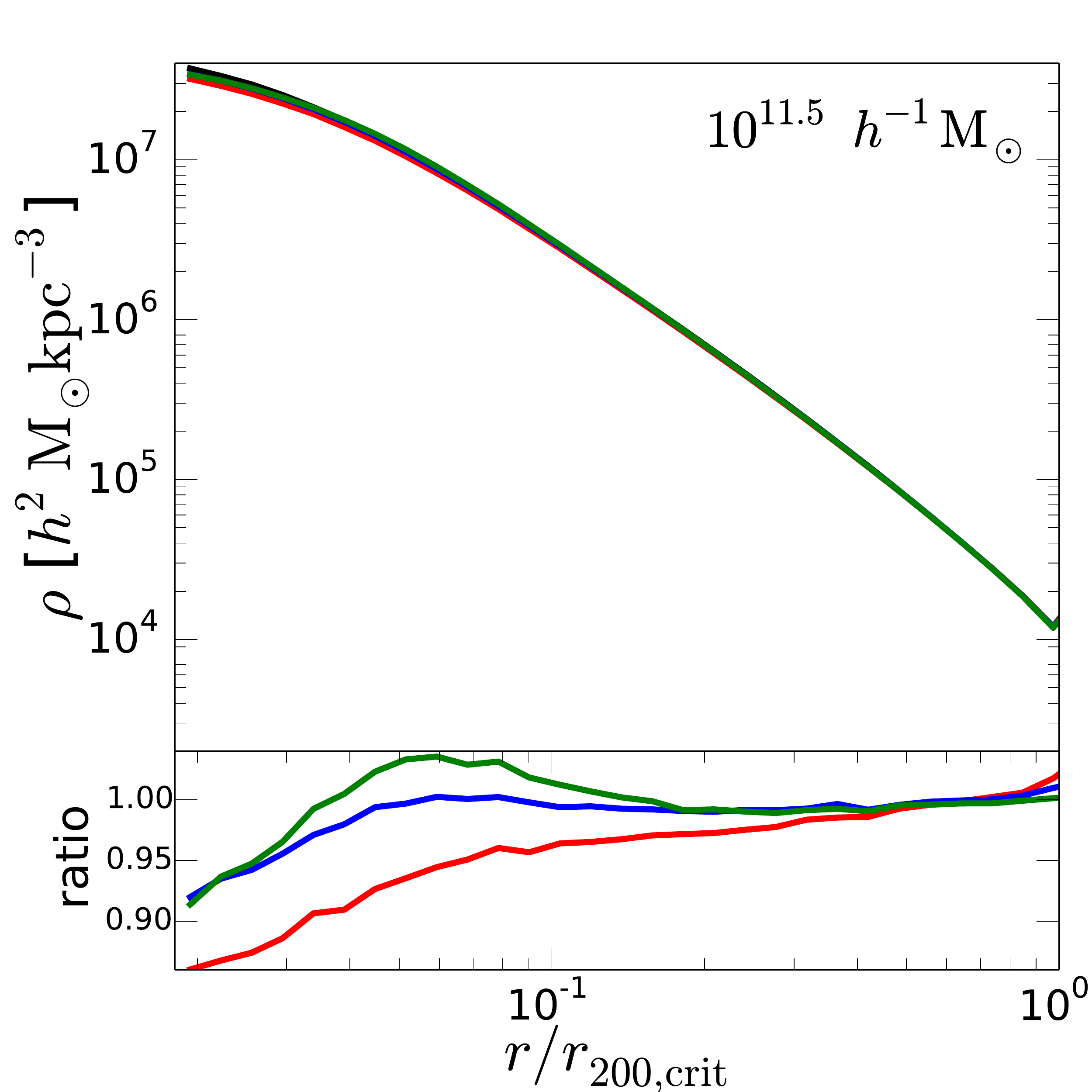}\\
\includegraphics[width=0.32\textwidth]{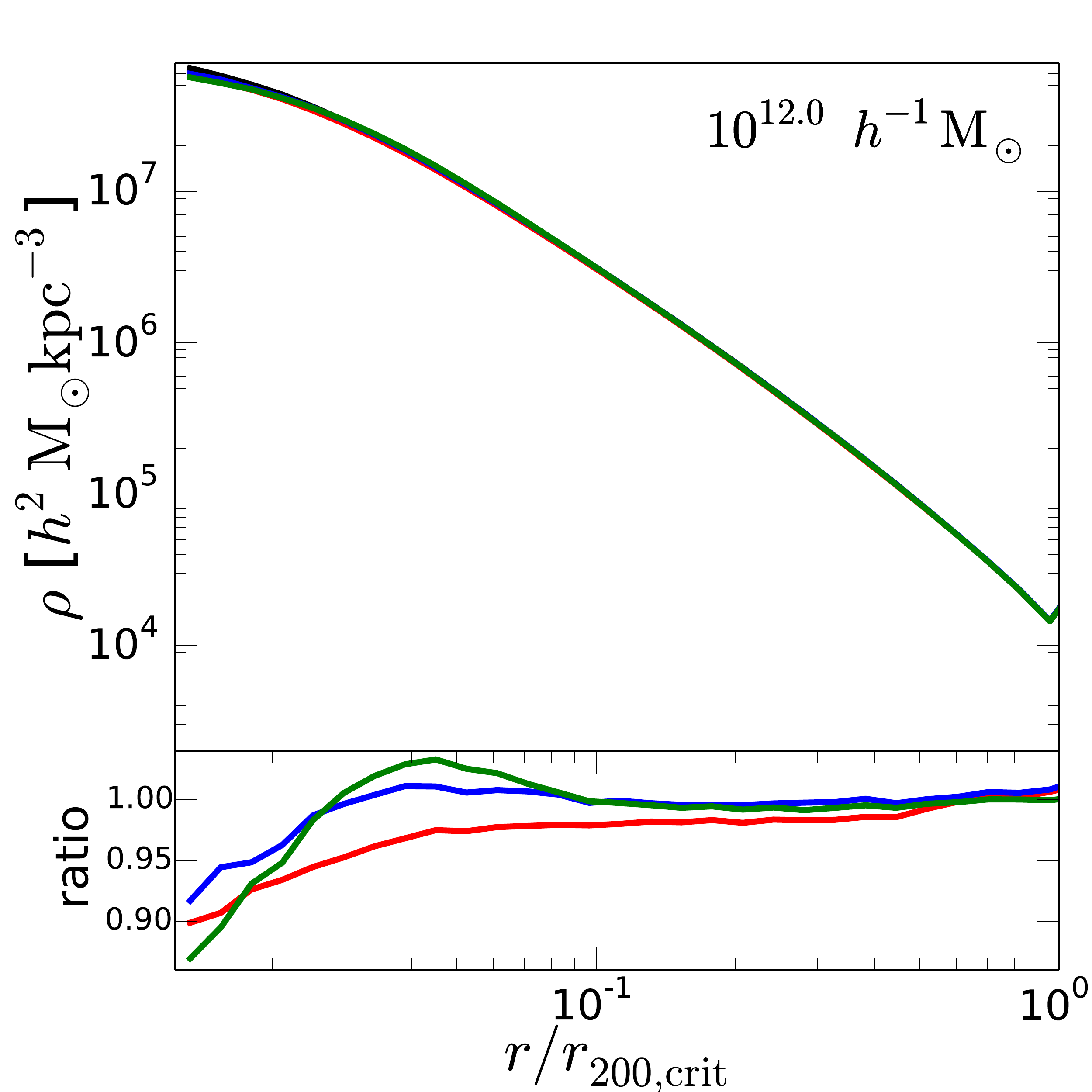}
\includegraphics[width=0.32\textwidth]{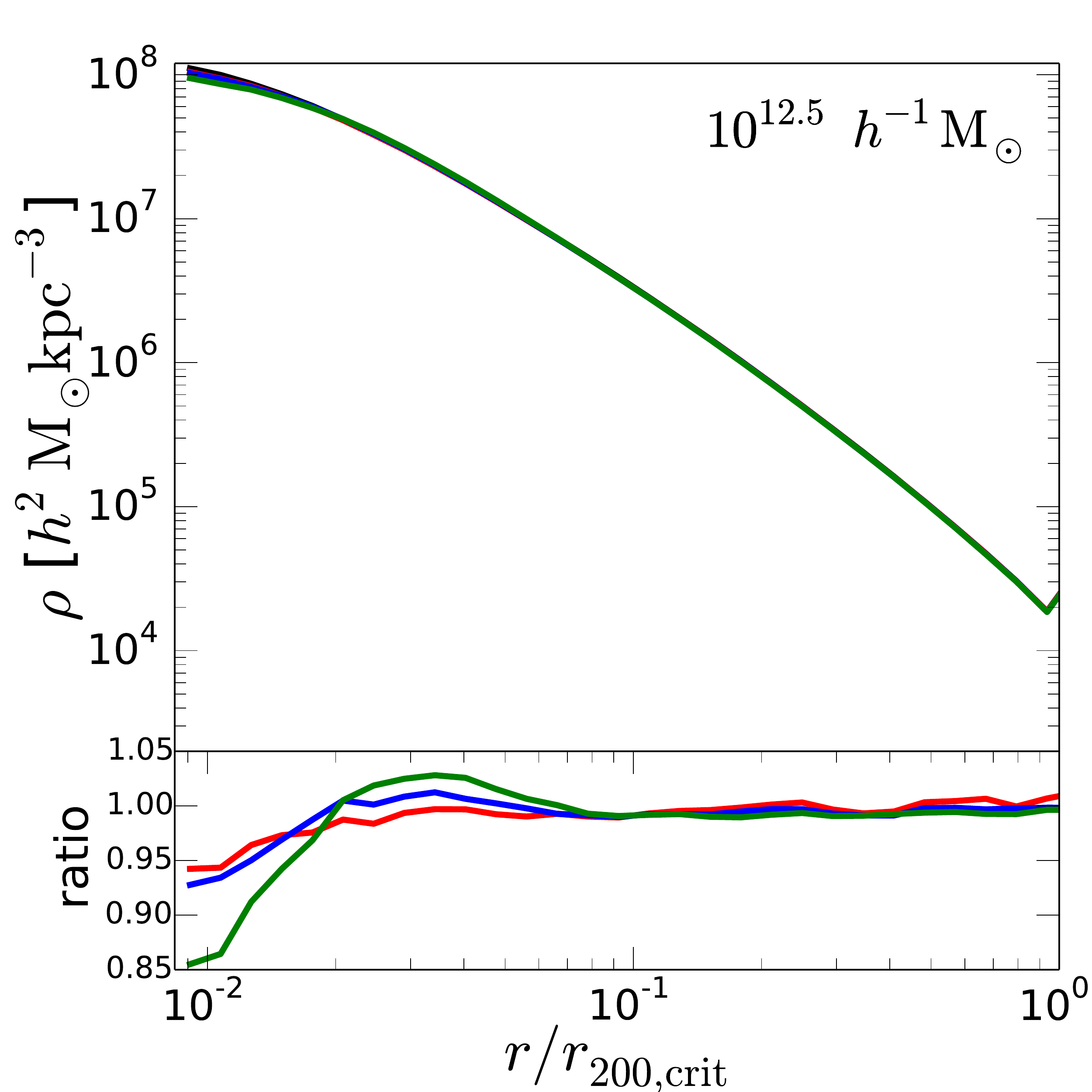}
\includegraphics[width=0.32\textwidth]{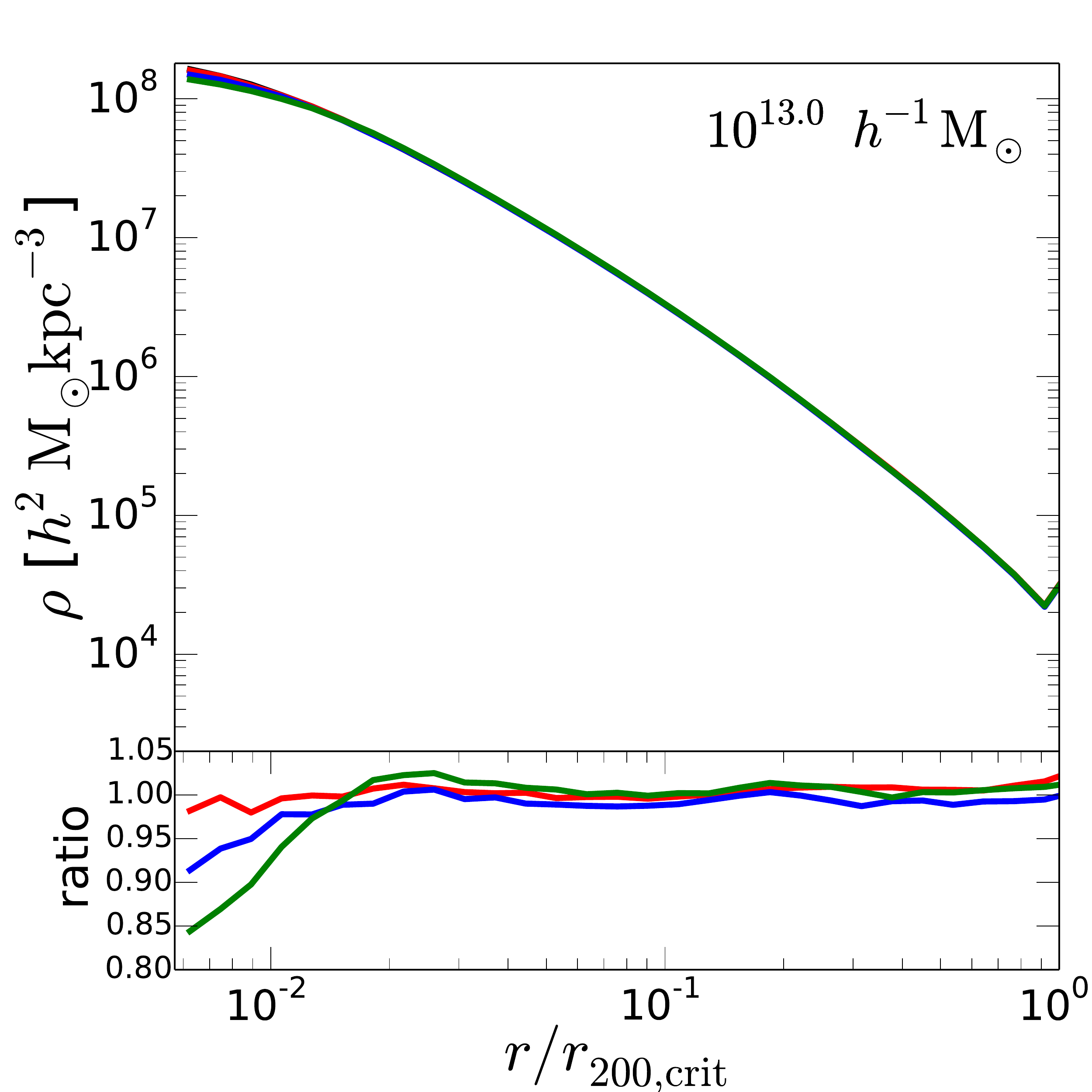}\\
\includegraphics[width=0.32\textwidth]{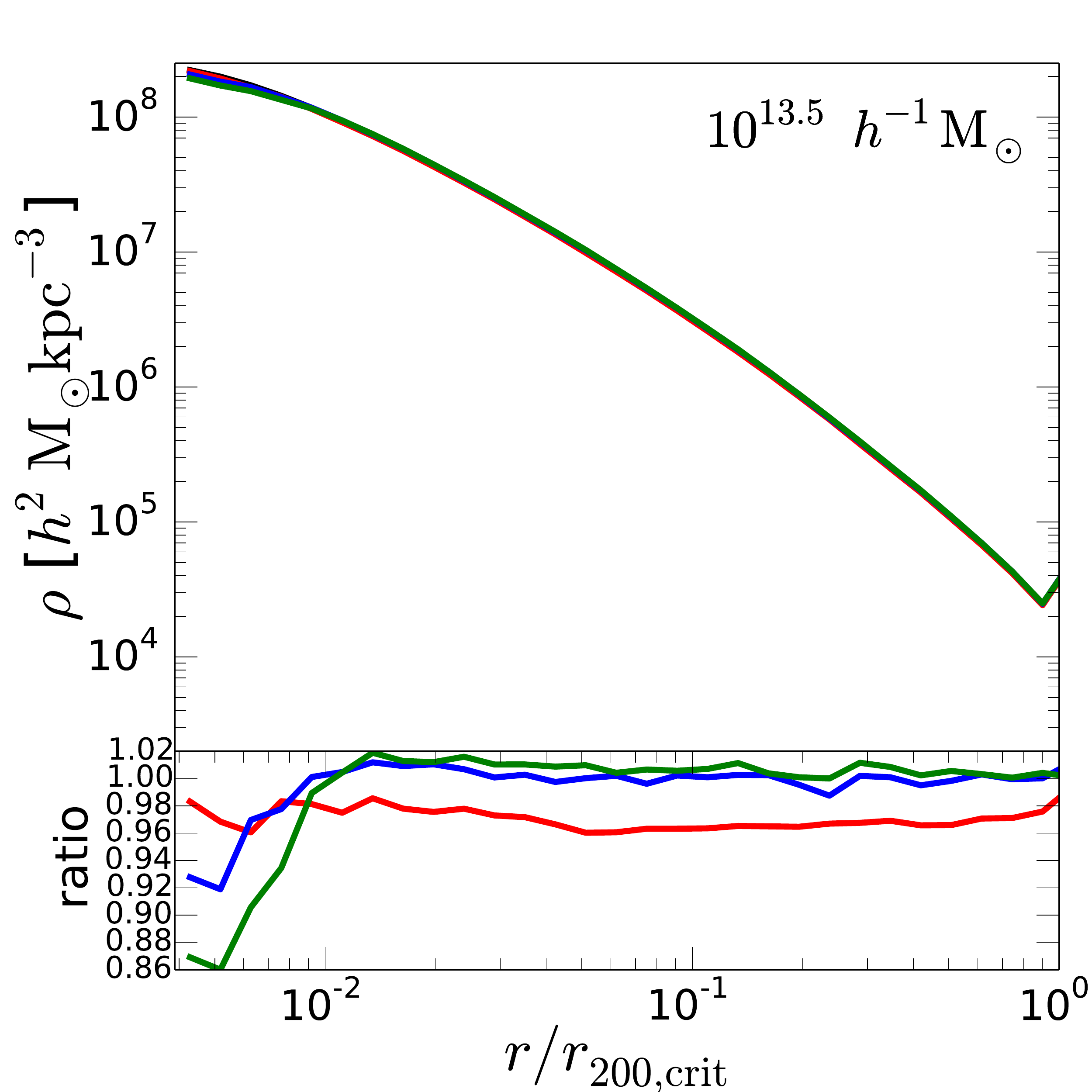}
\includegraphics[width=0.32\textwidth]{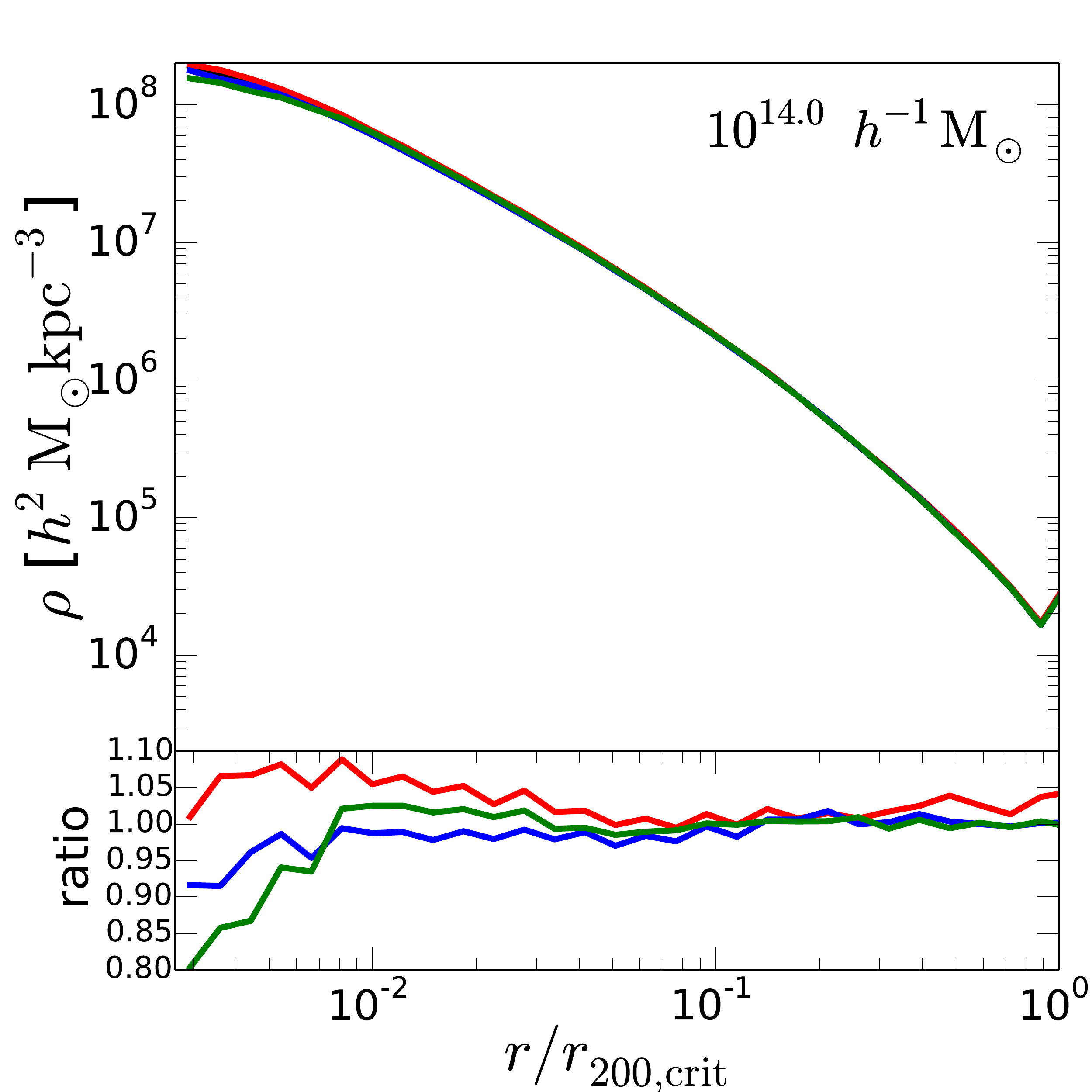}
\includegraphics[width=0.32\textwidth]{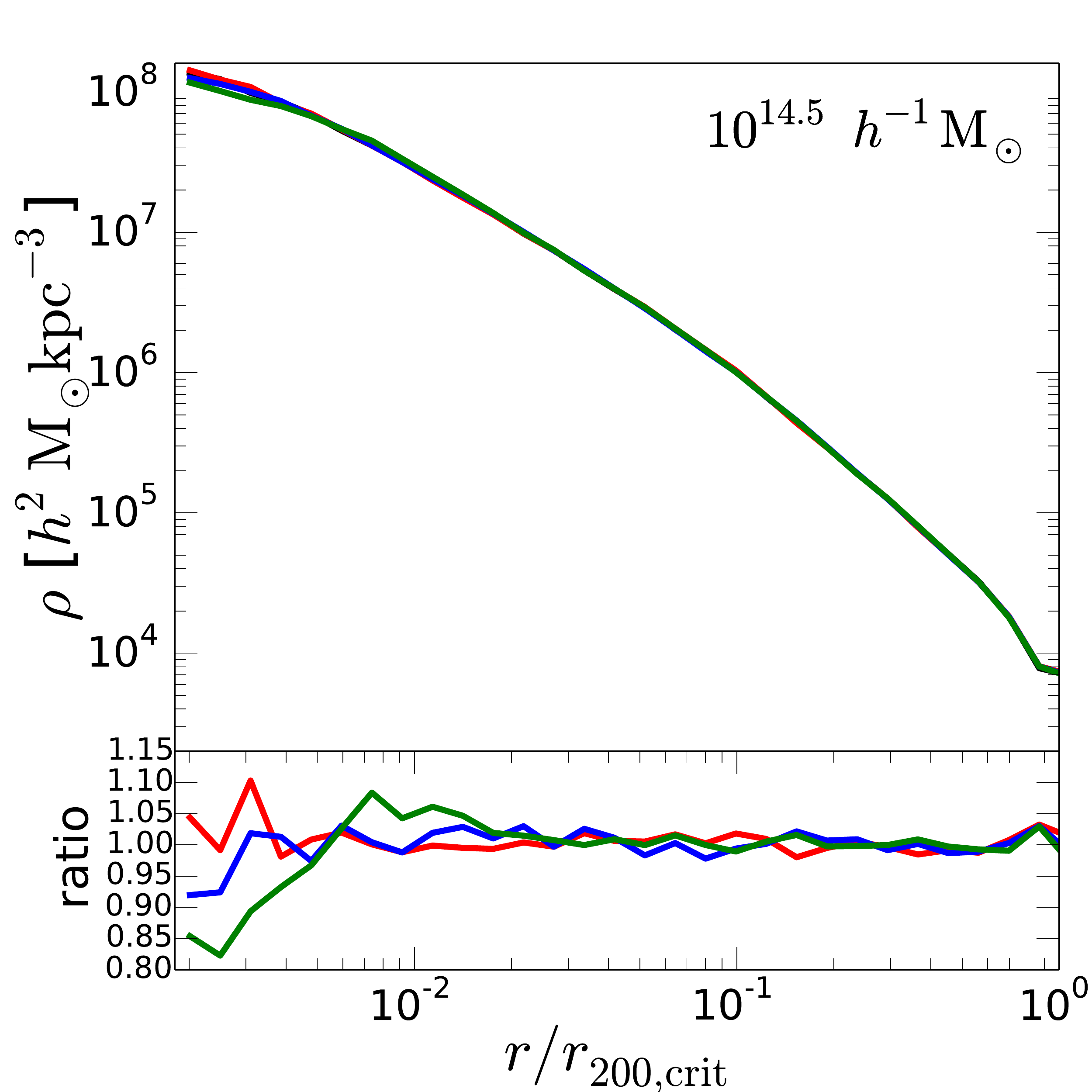}
\caption{Stacked density profiles for different halo mass ranges ($M_{\rm
200,crit}$) as indicated in each panel for our different DM models. We show the
profiles starting at $2\kpc$ out to the virial radius. One can clearly see that
the different non-CDM models affect the profiles in rather different ways
depending on the mass scale.}
\label{fig:density_profiles}
\end{figure*}

Although the power spectra are similar at $z=0$ between the different DM models,
there are significant differences in the halo mass function today due to the
delay in the formation of low mass haloes at high redshift.  This is shown in
Fig.~\ref{fig:massfct} where we plot the differential friend-of-friends (FoF) mass function at
$z=0$. Here we see a clear suppression of low mass haloes in ETHOS-1 to ETHOS-3 compared to
the CDM case (below a few times $\sim10^{11}\msun$ for model ETHOS-1). The strongest
suppression is seen for ETHOS-1 and the weakest for ETHOS-3. This is again expected given
the initial power spectra of the different models.  The lower panel of
Fig.~\ref{fig:massfct} shows that the suppression factor for haloes around
$\sim 10^{10}\hmsun$ is more than a factor of $2$ for ETHOS-1, whereas it is only about
$10\%$ for ETHOS-3, which is therefore quite close to the CDM case at this
dwarf-size scale.  We show below that the suppression of the faint end of
the halo mass function also carries over to the subhalo mass function. As
expected, the cutoff in the initial power spectra of ETHOS-1 to ETHOS-3 reduces the halo
abundance similarly as in WDM models. However, we stress again that the shape
of the power spectrum cutoff in our models is different from those in typical WDM models.

\begin{figure*}
\centering
\includegraphics[width=0.475\textwidth]{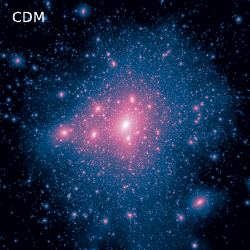}
\includegraphics[width=0.475\textwidth]{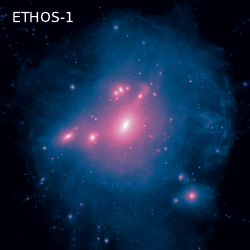}
\\\hspace{0.07cm}\includegraphics[width=0.475\textwidth]{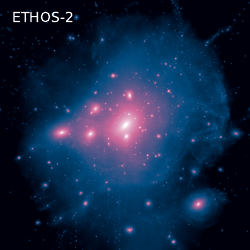}
\includegraphics[width=0.475\textwidth]{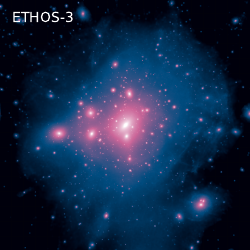}
\caption{DM density projections of the zoom MW-like halo simulations for four
different DM models. The suppression of substructure, relative to the CDM
model, is evident for the ETHOS models ETHOS-1 to ETHOS-3, which have a primordial power
spectrum suppressed at small scales. The projection has a side length and depth
of $500\kpc$.}
\label{fig:dm_density_small}
\end{figure*}

We attempt to analytically understand this cutoff by modeling the first DAO
feature in the linear power spectrum with a sharp power-law cutoff (see Eq.~\ref{eq:Mf}) and neglect the power contained in higher-order harmonics. Following the
derivation of \cite{vandenAarssen2012}, we adopt a
cutoff mass $M_\rmn{cut} = f M_\rmn{f}$ that scales linearly with the filtering
mass, $M_\rmn{f}$. The scaling factor, $f$, accounts for differences in halo
definitions (spherical overdensity vs. FoF) and for differences between the
simplified model of spherical collapse and our numerical cosmological
simulations.  We chose to adopt an exponential cutoff, $\exp(-M_\rmn{cut}/M)$,
to the differential CDM mass function of our particular realisation so that we
can exclude cosmic variance and halo suppression effects due to finite
resolution. This cutoff shape appears to accurately describe the abundance of
halos seeded by primordial power rather than artificial (numerical) power due to
particle discreteness effects~\citep[][]{Wang2007,
Angulo2013}\footnote{We note that according to the study by \citealt{Wang2007}, the limiting mass
below which discreteness effects are important is given by $M_{\rm lim}=10.1\Omega_m\rho_{\rm crit}dk_{\rm peak}^{-2}$, where
$d$ is the interparticle separation, and $k_{\rm peak}$ is the wave number for which $\Delta_{\rm linear}^2$ peaks. Taking the ETHOS-1 model, which
has the strongest power spectrum cutoff, we find that for our
parent $100h^{-1}$Mpc simulation, we have $M_{\rm lim}\sim1.7\times10^9h^{-1}$M$_\odot$, while
for the level-3 zoom simulations (the lowest resolution level for our zoom simulations), we have 
$M_{\rm lim}\sim4.2\times10^8h^{-1}$M$_\odot$. Our simulations are thus free of spurious artifacts in the mass 
scales we explored.}. 

Calibrating the global scaling factor, $f\approx2.85$,
we obtain a good match to the simulated FoF mass functions and find
\begin{equation}
  \label{eq:Mf}
  M_\rmn{cut} = 10^{11} \left(\frac{m_\rmn{WDM}}{\rmn{keV}}\right)^{-4} h^{-1}\rmn{M}_\odot.
\end{equation}
Alternatively, we can express the cutoff as a function of the kinetic decoupling temperature
   $T_\mathrm{kd}$. In this case, we find
 \begin{equation}
    M_\mathrm{cut}= 5\times 10^{10} \left(\frac{T_\mathrm{kd}}{100\,\rmn{eV}}\right)^{-3} h^{-1} \rmn{M}_\odot\,.
  \end{equation}
This result for the evolved, \emph{non-linear} power spectrum agrees to within about a 
factor of 2 with earlier analytic estimates for this 
regime~\citep[][]{Loeb2005, Bertschinger2006, Bringmann2009}, based on definitions of $M_\mathrm{cut}$ 
related to the form of the \emph{linear} power spectrum. Note that the above relation between 
$M_\mathrm{cut}$ and $T_\mathrm{kd}$ specifically assumes the definition of $T_\mathrm{kd}$ proposed 
in \cite{Bringmann:2006mu}; other definitions of $T_\mathrm{kd}$ would change it by a constant 
factor \cite{Bringmann:2016ilk}.

We also observe indications for a slight deviation from the $T_\mathrm{kd}^{-3}$ dependence;
fully resolving this would however require to run high-resolution simulations over a much larger
range of cutoff values, which is beyond the scope of this work.
As a result, the effect of early DM-DR interactions on the mass function (around
the cutoff region) can be approximated by an exponential cutoff due to free
streaming of WDM (see yellow dashed lines in Fig.~\ref{fig:massfct}). However, this simple approximation ceases to be
valid at masses much smaller than $M_\rmn{cut}$ because SIDM models have more
power in in the primordial power spectrum in comparison to WDM, which is
characterised by a comparably sharp cutoff (as explicitly demonstrated by the
discrepancies of the mass functions of the analytical model and the ETHOS-1
simulation). In the non-linear regime of structure formation, adjacent-modes do
not evolve independently of each other and start to couple redistributing
power to the valleys at the expense of the extrema. Hence, the shallower decay
towards small halos is seeded by the integrated additional power of SIDM models
in comparison to WDM.

The effects we have discussed so far are mainly driven by the primordial
damping of the DM power spectrum. They are not strongly affected by late time
DM self-interactions.  As we know, self-interactions will affect the internal
structure of haloes at late times, where the density is high enough to cause at
least some particle collisions during a Hubble time. We can try to quantify
this already at the resolution level that our parent simulation allows. To do
this, we measure the central or core density for all resolved main haloes in
the uniform box simulations, similar to the analysis presented in
\cite{Buckley2014}.  The mass resolution of our uniform box is slightly better
than that of \cite{Buckley2014}, and we probe at the same time a volume which
is about $3.8$ times larger. We can therefore sample a larger range of halo masses and
with better statistics.  We define the central (core) density within 
three times the softening length ($8.7\kpc$). The upper panel of
Fig.~\ref{fig:cores} shows the actual core density, while the lower panel shows
the ratio with respect to the CDM case. We take the median value of the
distribution within each mass bin. The
plot shows the familiar scale of density with mass at a fixed radius, with core
densities that vary from $\sim 10^6\,h^2{\rm M}_\odot {\rm kpc}^{-3}$ for halo
masses around $\sim 10^{10}\hmsun$ to $\sim 10^8\,h^2{\rm M}_\odot {\rm
kpc}^{-3}$ for halo masses around $\sim 10^{14}\hmsun$.  Models ETHOS-1 (red) and ETHOS-2 
(blue) have a significantly reduced core density compared to the CDM case for
low mass haloes. We note that the effect is strongest in the former than in
the latter, which points to the primordial power spectrum suppression as the
main culprit since the cross section is lower for model ETHOS-1 than for model ETHOS-2.
Low-mass haloes in ETHOS-1 are therefore less dense than in CDM, mainly because they
form later (analogous to the WDM case). Interestingly, ETHOS-3 shows a different
behaviour. Here the core density is most reduced for massive systems. This
effect is also slightly seen for ETHOS-2, but it is much more pronounced for ETHOS-3.
The cause of this behaviour is DM self-scattering since ETHOS-3 has a large
cross section, even at large scales, thus, an impact on the inner structure of
haloes is expected even for massive haloes (see Fig.~\ref{fig:models}).  The
reason we do not see a reduction in $\rho(3\epsilon)$ at low masses for the ETHOS-3 
case, despite its very large cross section at these scales, is because
$3\epsilon\sim9$~kpc is  too large relative to the maximum core sizes (set by
$\lesssim r_{-2}$, the radius where the logarithmic density slope is $-2$) 
that low mass haloes ($<10^{11}\hmsun$) can have.  

To better understand the core density behaviour we show in
Fig.~\ref{fig:density_profiles} stacked radial density profiles splitting the
main haloes in different mass bins (as indicated in the legends). One can
clearly see that the haloes at the low mass end are significantly affected by
both the damping of the initial power spectrum and self-interactions. The
strongest density reduction occurs here for ETHOS-1, which has the largest damping
scale. This is true despite model ETHOS-1 having the lowest self-interaction of all
models. In other words, for the models studied here (ETHOS-1 to ETHOS-3), the suppression of the
power spectrum at small scales dominates over the effect of self-interactions
at low masses. This trend continues up to MW masses, where a reversal of the
hierarchy of models begins to happen, with ETHOS-3 becoming the most divergent
relative to CDM.  Massive haloes are not strongly affected by the damping of
primordial fluctuations making self-interactions the dominant mechanism. The
reduction of the central density therefore responds directly to the amplitude
of the cross section. 

\subsection{Galactic halo}

\begin{figure}
\centering
\includegraphics[width=0.475\textwidth]{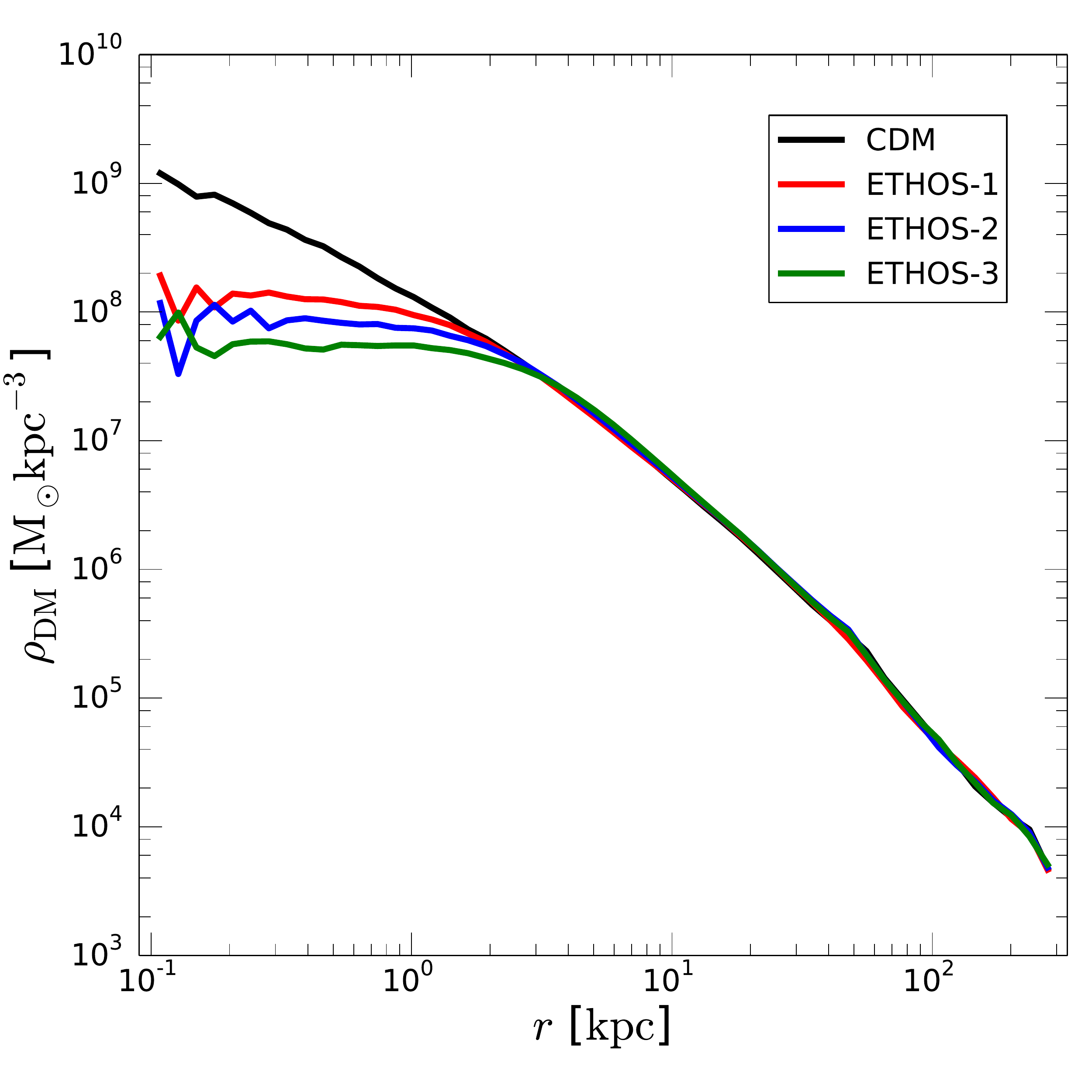}
\caption{Spherically averaged density profiles of the Milky-Way-size haloes for
the four different DM models. The non-CDM models have only a mild impact on the
central density profile where self-interactions lead to the formation of a
small core ($\sim 2\kpc$), with a size that correlates with the amplitude of the
scattering cross section. The damping in the initial power spectrum for these models
(see left panel of Fig.~\ref{fig:models}) has only a
secondary impact on the density profiles of MW-size haloes.}
\label{fig:main_profile}
\end{figure}

\begin{figure*}
\centering
\includegraphics[width=0.475\textwidth]{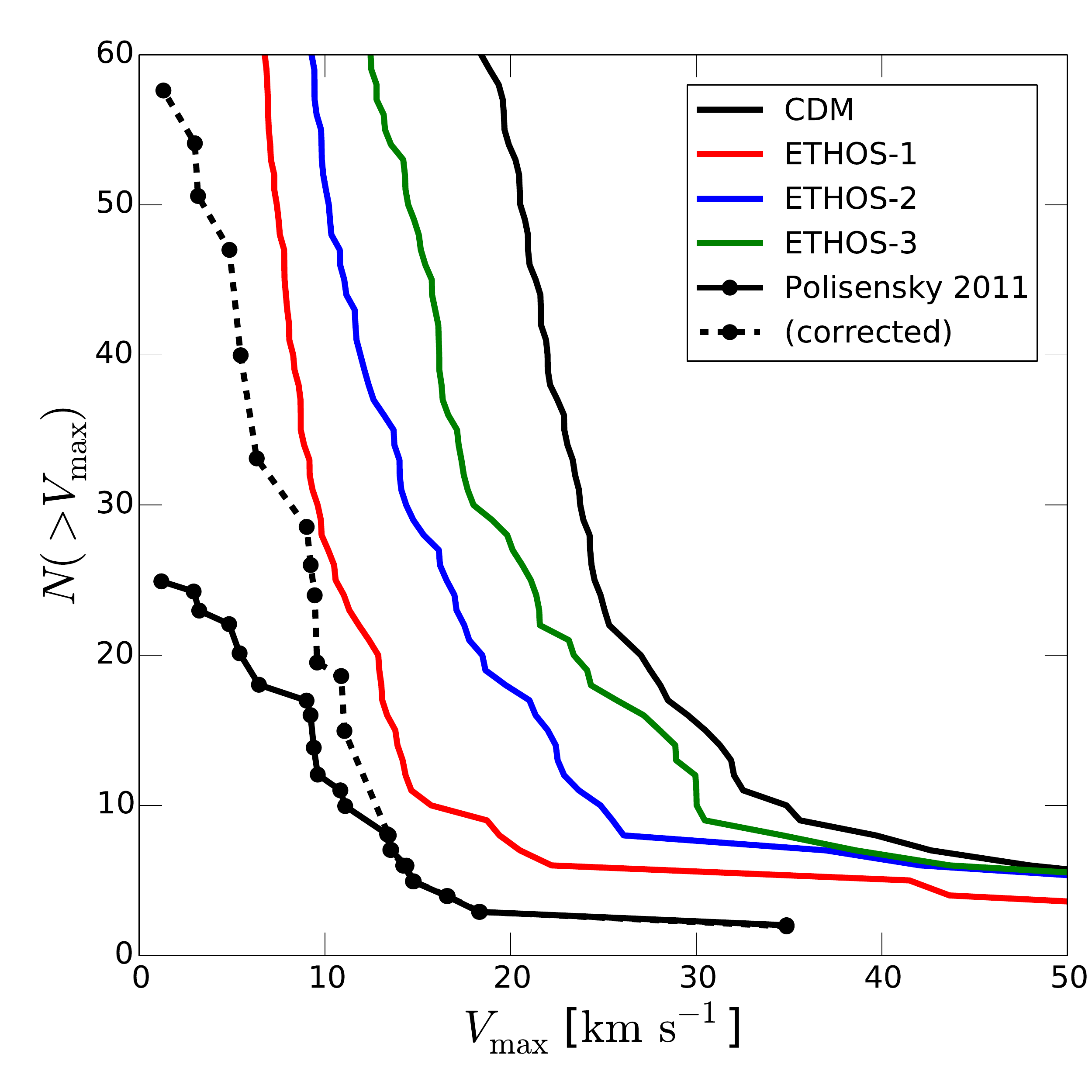}
\includegraphics[width=0.475\textwidth]{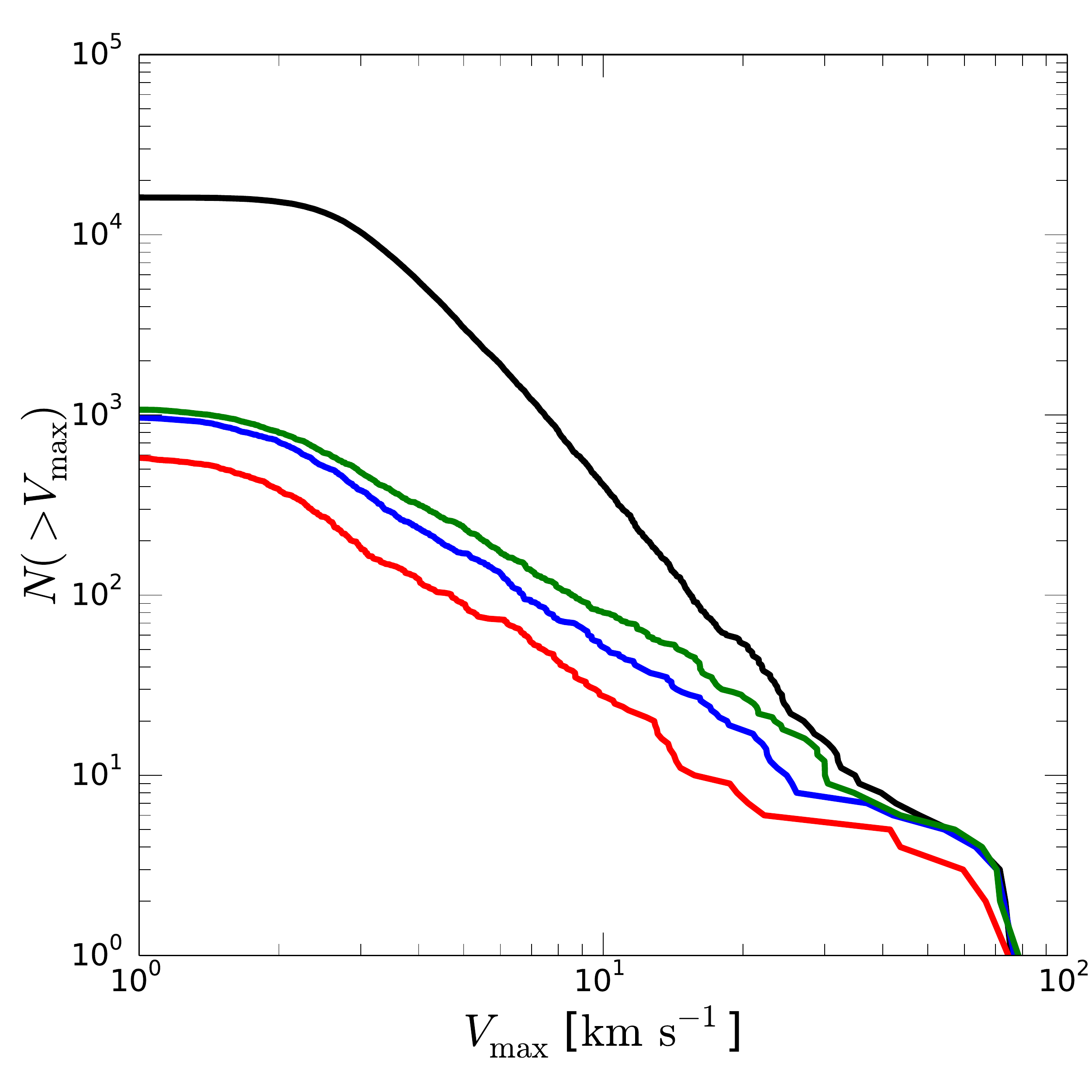}
\caption{The number of subhaloes as a function of their maximal circular
velocity for the four different DM models. We include all subhaloes with a
halocentric distance less than $300\kpc$. Left panel: linear scale with a
comparison to observed satellites of the Milky Way including a sky coverage correction~\citep{Polisensky2011}. The
discrepancy between the number of observed satellites and resolved DM subhaloes
is significantly reduced in the models ETHOS-1 to ETHOS-3. Right panel: log-log
scale. The plateau at low $V_{\rm max}$ values is caused by limited resolution.}
\label{fig:sub_vmax}
\end{figure*}

We will now focus on the analysis of the zoom-in simulation of the selected
MW-sized halo. We first discuss the models ETHOS-1 to ETHOS-3 and later focus on ETHOS-4, which
addresses some of the small-scale issues of CDM.
Fig.~\ref{fig:dm_density_small} shows the projected DM density distribution on
the scales of the MW halo for the CDM case, and for models ETHOS-1 to ETHOS-3\footnote{high resolution images and movies can be found at \url{http://mvogelsb.scripts.mit.edu/ethos.php}}.  At these
scales ($\lesssim500$~kpc), the suppression of small scale structure is clearly
visible. This suppression is driven by the resolved cutoff scale in the linear
power spectra of ETHOS-1 to ETHOS-3 caused by DM-DR interactions.  This cutoff reduces the
number of resolved subhaloes very strongly for model ETHOS-1, which has the largest
damping scale. We stress once more, that self-interactions of the order
discussed here mainly affect the internal structure of haloes, without significantly
altering the abundance of subhaloes within MW-like haloes.
Fig.~\ref{fig:dm_density_small} also demonstrates that the position and
appearance of the largest subhaloes does not change significantly across the
different models (except perhaps for ETHOS-1).

Due to the amplitude and velocity dependence of the cross section for models
ETHOS-1 to ETHOS-3,  there is only a rather mild impact of particle collisions on the main
halo properties.  This can be seen in Fig.~\ref{fig:main_profile}, where we
show the density profile of the main MW halo. There is a resolved DM core with
a size of $\sim 2\kpc$, but it is clearly much less pronounced than the effect
seen for a large constant cross section of $\sim10\cpm$, where the core size is
$\gtrsim50\kpc$~\citep[e.g.][]{Vogelsberger2012}, this is naturally expected
since at the characteristic velocities of DM particles in the MW halo, the
cross sections in our models are $<1\cpm$ (see Fig.~\ref{fig:models}). Such
small cores in the MW galaxy (or Andromeda) are fully consistent with any
constraints from observations \citep[e.g.][]{Bovy2013}.
Fig.~\ref{fig:main_profile} also demonstrates that the core is largest for
model ETHOS-3 and smallest for model ETHOS-1, which is directly explained by the relative
amplitude of the cross section in these models.  We note that at the scales of
the MW halo, the cutoff scale in the linear power spectrum (left panel of
Fig.~\ref{fig:models}) has a subdominant impact compared to the effect of DM
collisions. This was already seen, albeit not as clearly, in
Fig.~\ref{fig:density_profiles}. 

The apparent reduction of substructure is quantified in more detail in
Fig.~\ref{fig:sub_vmax}, where we show the cumulative distribution of subhaloes
within $300\kpc$ of the halo centre as a function of their peak circular
velocity $V_{\rm max}$.  The left panel shows the cumulative number on a linear
scale, and includes observational data from \cite{Polisensky2011}. The 
MS problem is apparent since there are significantly more CDM subhaloes
than visible satellites. This discrepancy can be solved or alleviated through a
combination of photo-evaporation and photo-heating when the Universe was reionised, and supernova
feedback~\citep[e.g.][]{Efstathiou1992,Gnedin2000,Benson2002, Koposov2008}, although photo-evaporation and photo-heating alone may not be enough to bring the predicted
number of massive, luminous satellites into agreement with observations
~\citep[e.g., ][]{Boylan2012, Brooks2013}.  The plot also demonstrates that the reduction of
substructure in ETHOS-1 to ETHOS-3 alleviates the abundance problem significantly. The
strong damping in the power spectrum of model ETHOS-1 leads to a very significant
reduction of satellites which is quite close to the data, perhaps too close
given the expected impact of reionisation and supernovae feedback. If these
processes were to be included in our simulations with a similar strength as
they are included in hydrodynamical simulations within CDM, model ETHOS-1 would be
ruled out. One must be cautious however, since the strength of these processes
is not known well enough, they could in fact be much weaker than currently
assumed within CDM, which would lead to different conclusions. Models ETHOS-2 and
ETHOS-3 show a smaller but still significant reduction of the abundance of
subhaloes which spans the range between the CDM result and the observed
satellite population.  The right panel of Fig.~\ref{fig:sub_vmax} shows the
same quantity on a logarithmic scale. We note that our main objective here is 
to show that in the ETHOS models we explored, there is a substantial effect on the
number of satellites (due to the power spectrum cutoff) and in the inner halo
densities (a combination of the power spectrum cutoff and the self-interaction
cross section). A more definitive solution to the MS problem within
ETHOS would require the inclusion of the baryonic processes mentioned above. It is
interesting to note that the reduction to the subhalo abundance in ETHOS-3 is 
approximately a factor of 2 at 20 km/s, which is only slightly smaller than the reduction
achieved by current hydrodynamical simulations that include baryonic physics (see Fig. 4 of \citealt{Sawala2014})

\begin{figure*}
\centering
\includegraphics[width=0.475\textwidth]{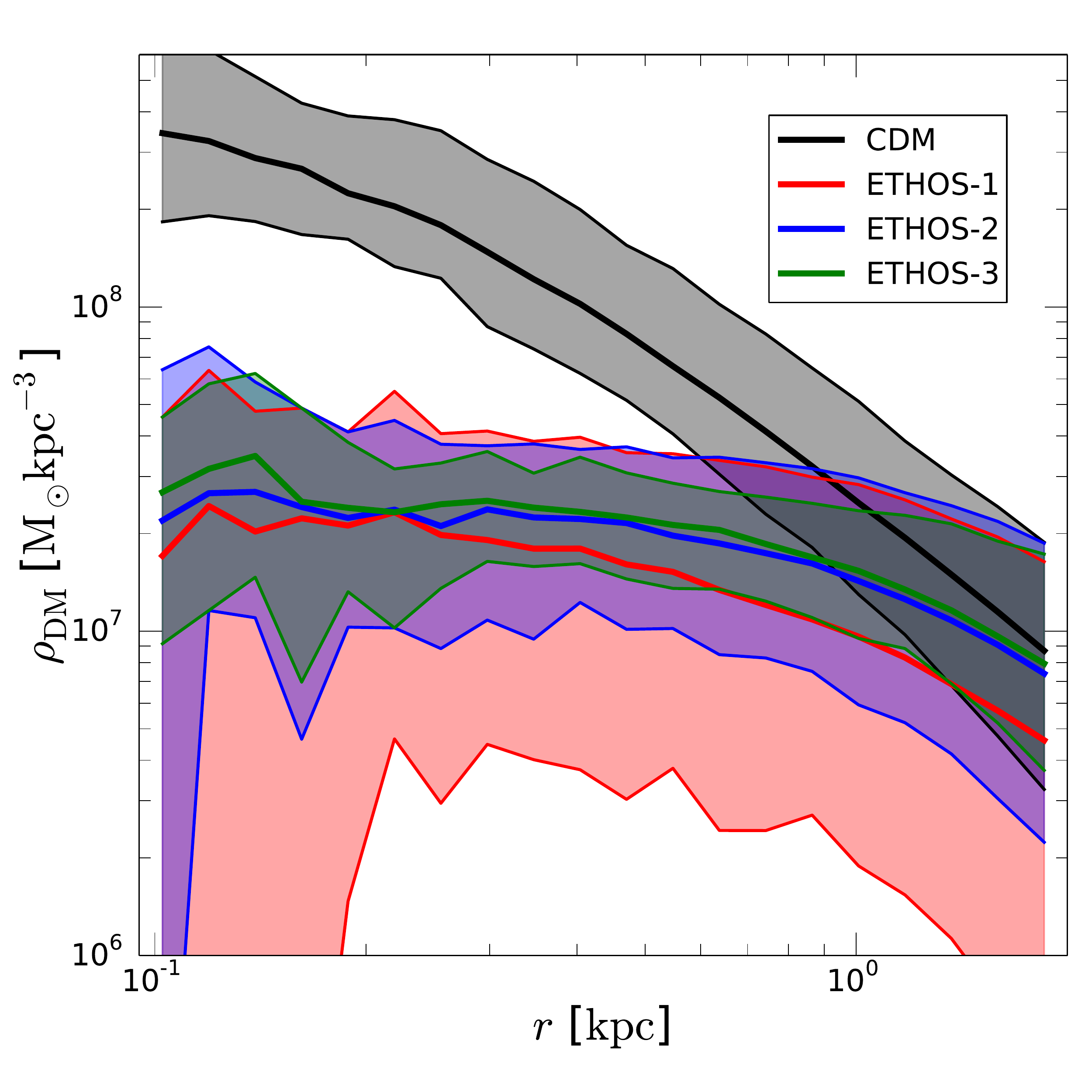}
\includegraphics[width=0.475\textwidth]{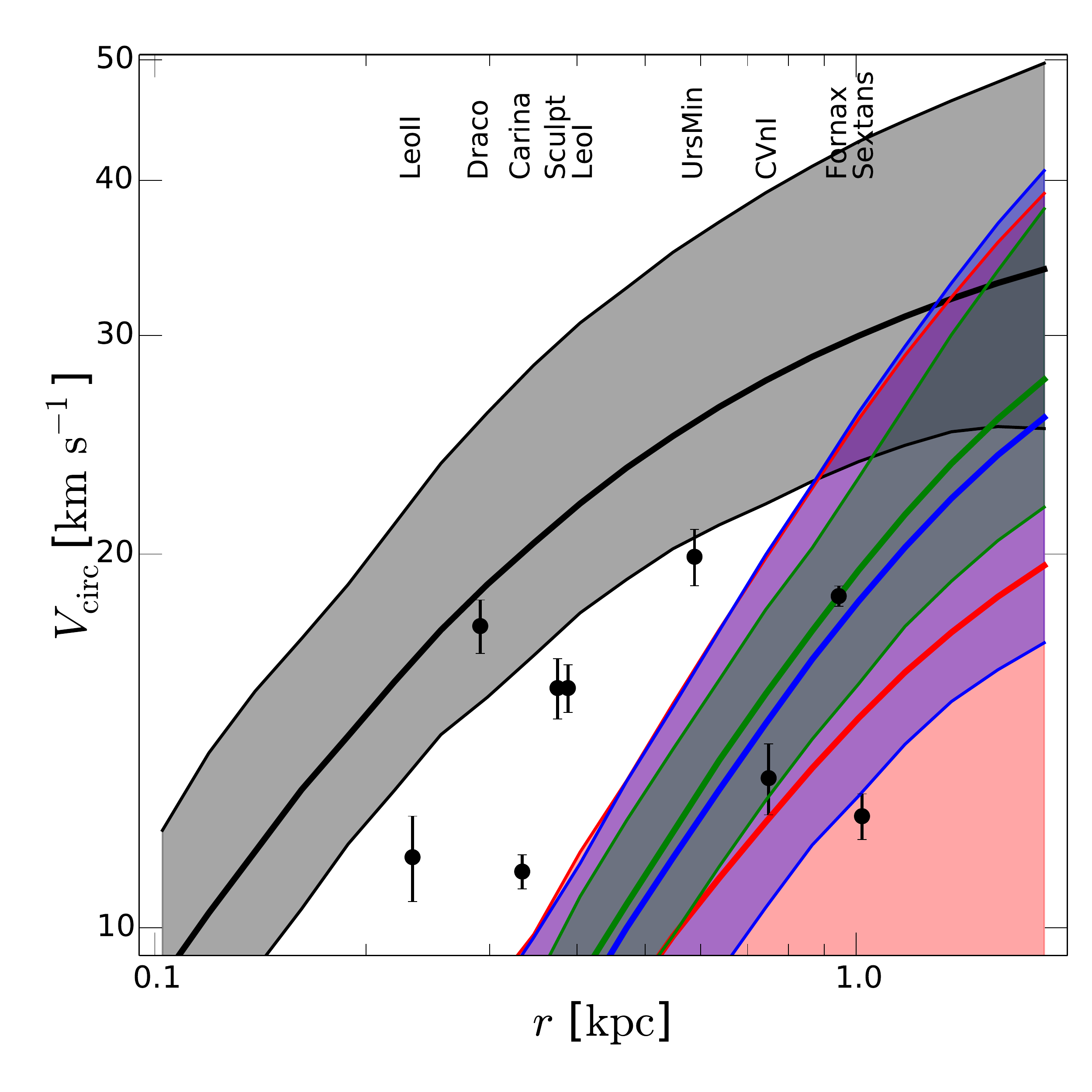}
\caption{Internal subhalo structure for different DM models. Left panel: Density profile of the $15$
subhaloes with the largest $V_{\rm max}$ within $300\kpc$. The CDM subhaloes have cuspy density profiles
while the subhaloes in models ETHOS-1 to ETHOS-3 develop a~$\gtrsim2\kpc$ core. Right panel:
Circular velocity profiles of the same subhaloes. The data points on the right are for the classical MW dSphs
taken from \protect\cite{Wolf2010}.  The CDM model shows the TBTF 
problem; i.e. the most massive subhaloes in the simulation are too dense to host most of the
classical MW dSphs. The combination of damping in the initial
power spectrum and self-interactions lead to a very drastic reduction of
inner densities and circular velocities (enclosed mass), which seem to be inconsistent 
with the data.}
\label{fig:sub_internal}
\end{figure*}

Finally, we show the (radially averaged) internal structure of subhaloes in
Fig.~\ref{fig:sub_internal}. Here we select for each model the $15$ subhaloes
with the largest present-day $V_{\rm max}$ values and plot the distribution of
density profiles (left panel) and circular velocity profiles (right panel).
The thick lines show the median of the distribution, whereas the shaded region
shows the lower and upper envelopes. We only consider subhaloes that are within
$300\kpc$ halocentric distance.   The right panel also contains observational
data from nine of the classical MW dSphs taken from~\cite{Wolf2010}.  The
distribution of circular velocities for the CDM model clearly shows the
well-known TBTF problem, while the density profiles points to the also
well-known CC problem: if Fornax is cored for instance, the size of its
DM core is between 0.6 and 1.8 kpc \citep{Amorisco2013}, which cannot be
accommodated in the CDM model. The non-CDM models on the other hand show
clearly reduced (density) circular velocity profiles, which are seemingly too low
compared to what is required by the data.  We show below that the
combination of self-interactions and the damping in the initial power spectrum
is responsible for this strong reduction. This is quite different from the
results shown in~\cite{Vogelsberger2012}, where we only considered the effect
of self-interactions.  The fact that models ETHOS-1 to ETHOS-3 do not fit the data is
unsurprising since those models were not specifically tuned to reproduce the
subhalo statistics of the MW. They were rather picked out of a large
particle physics parameter space. This result however shows that it can
actually be quite difficult to predict a priori the combined impact of
self-scattering and primordial damping on the highly non-linear evolution that
leads to the internal structure of subhaloes. It also shows the potential of
the MW satellites data to constrain the parameter space of our effective framework.

We should note that the severity of the 
TBTF problem in the MW depends on the MW halo mass. For low halo masses 
$\lesssim8\times10^{11}$M$_{\odot}$, the problem disappears \citep[e.g.][]{Wang2012}.
Since we are using a halo with a mass of $1.6\times10^{12}$M$_{\odot}$, the TBTF is
quite clear in our CDM simulation. This dependence on the halo mass has been discussed
elsewhere. For the purpose of our work, we show a case where the problem is clear,
and focus on studying the DM models that can alleviate it.

\subsection{Disentangling the impact of late DM self-interactions versus early DM-DR interactions}

\begin{figure*}
\centering
\includegraphics[width=0.475\textwidth]{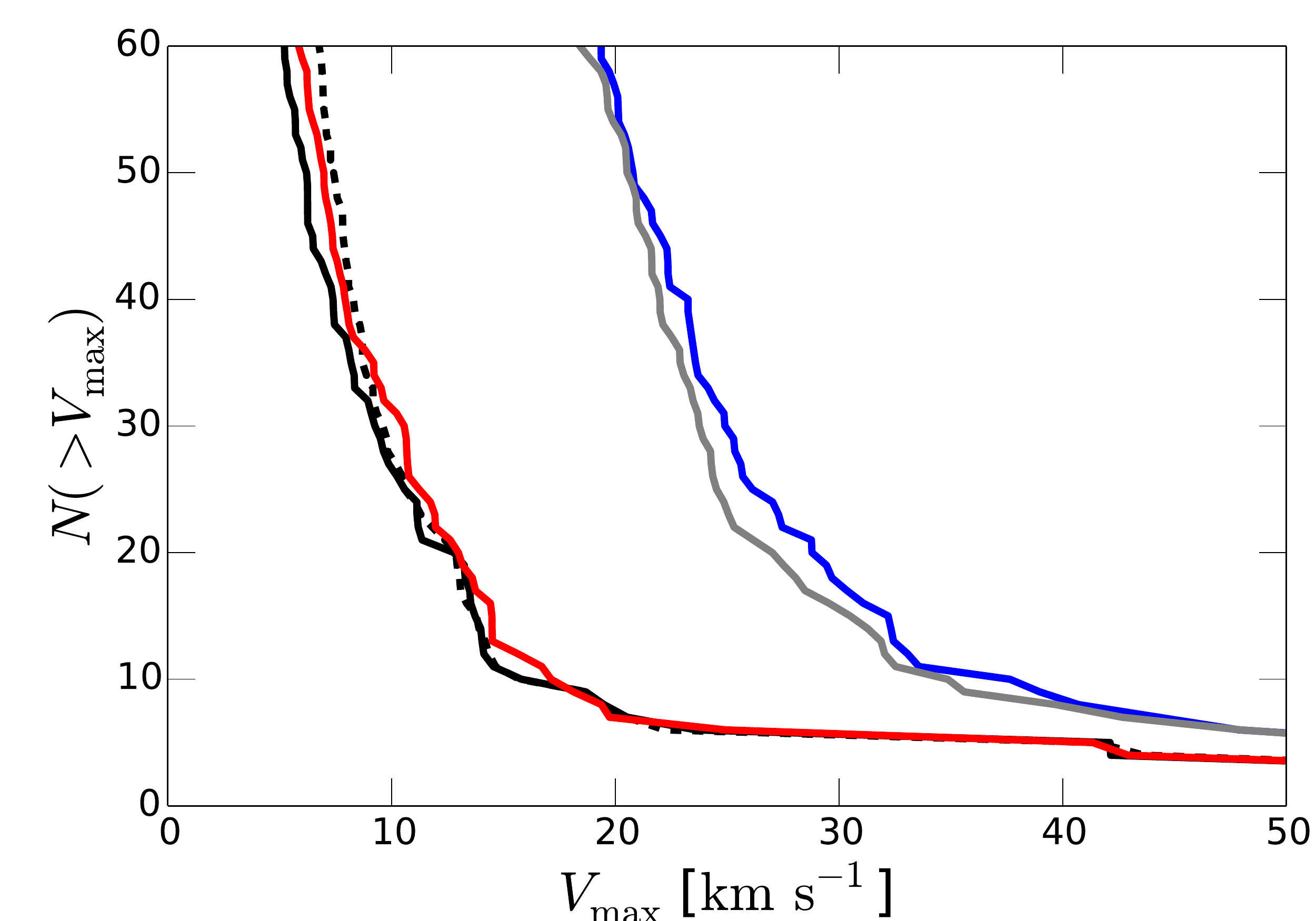}
\includegraphics[width=0.475\textwidth]{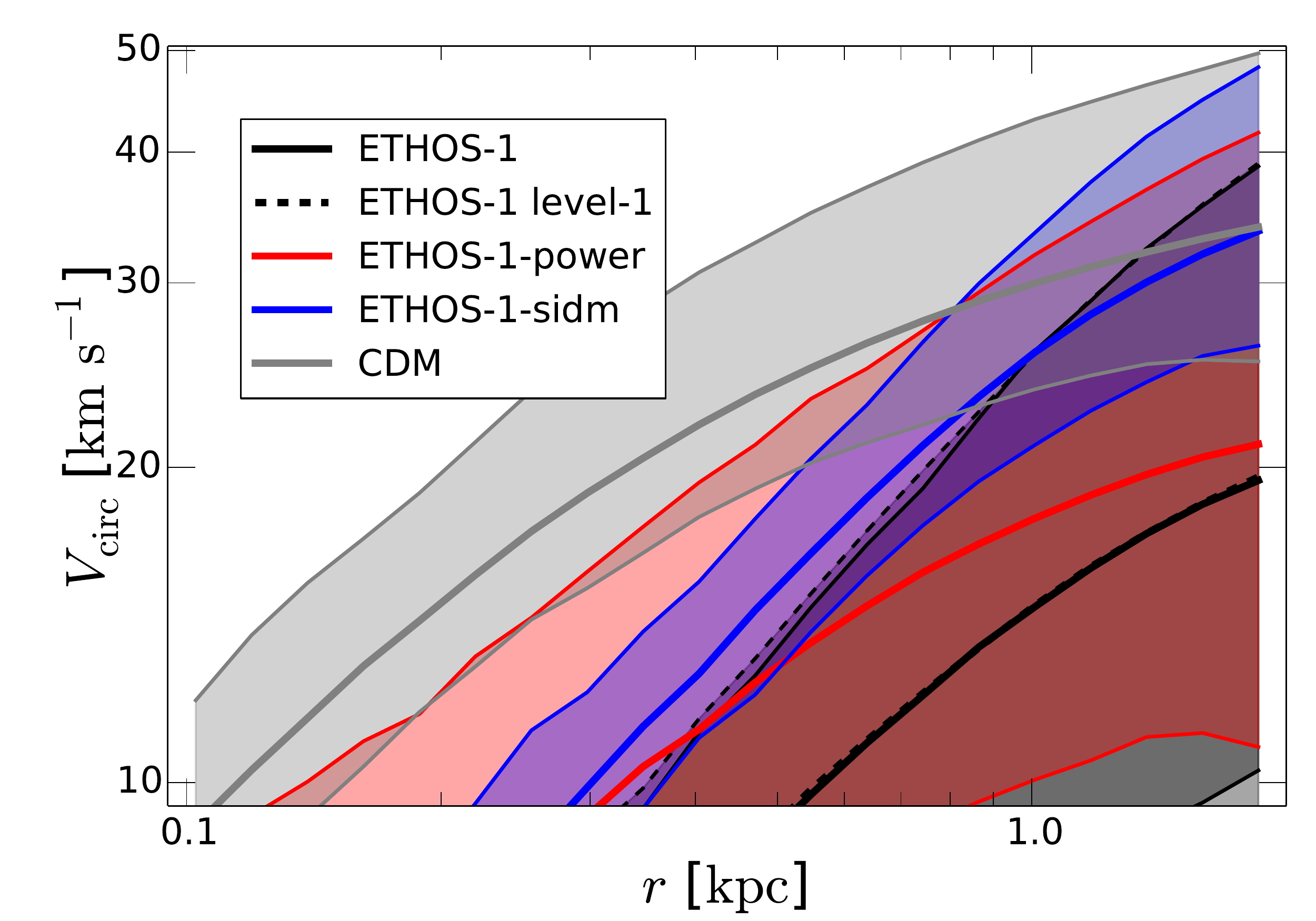}\\
\vspace{-0.5cm}
\includegraphics[width=0.475\textwidth]{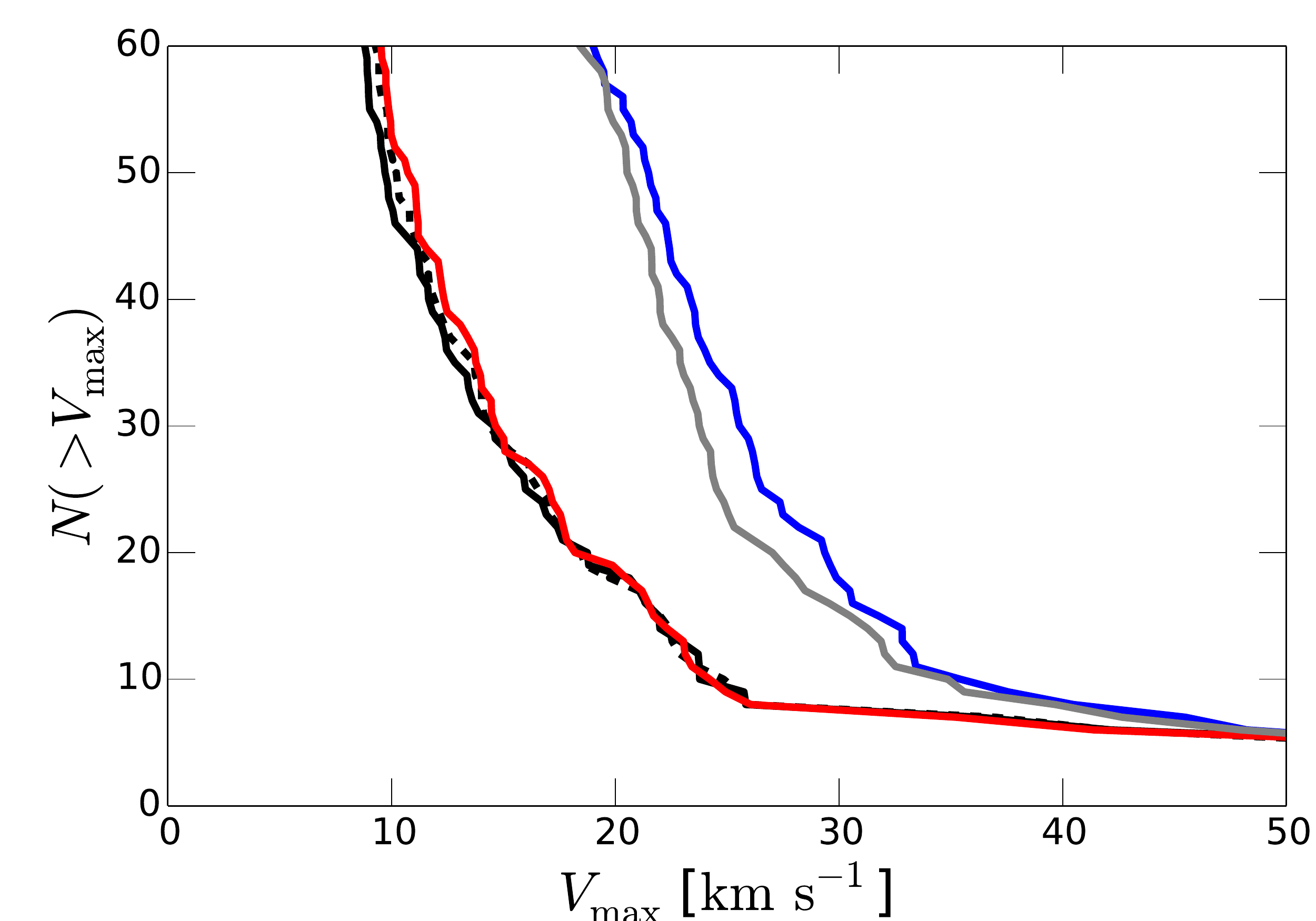}
\includegraphics[width=0.475\textwidth]{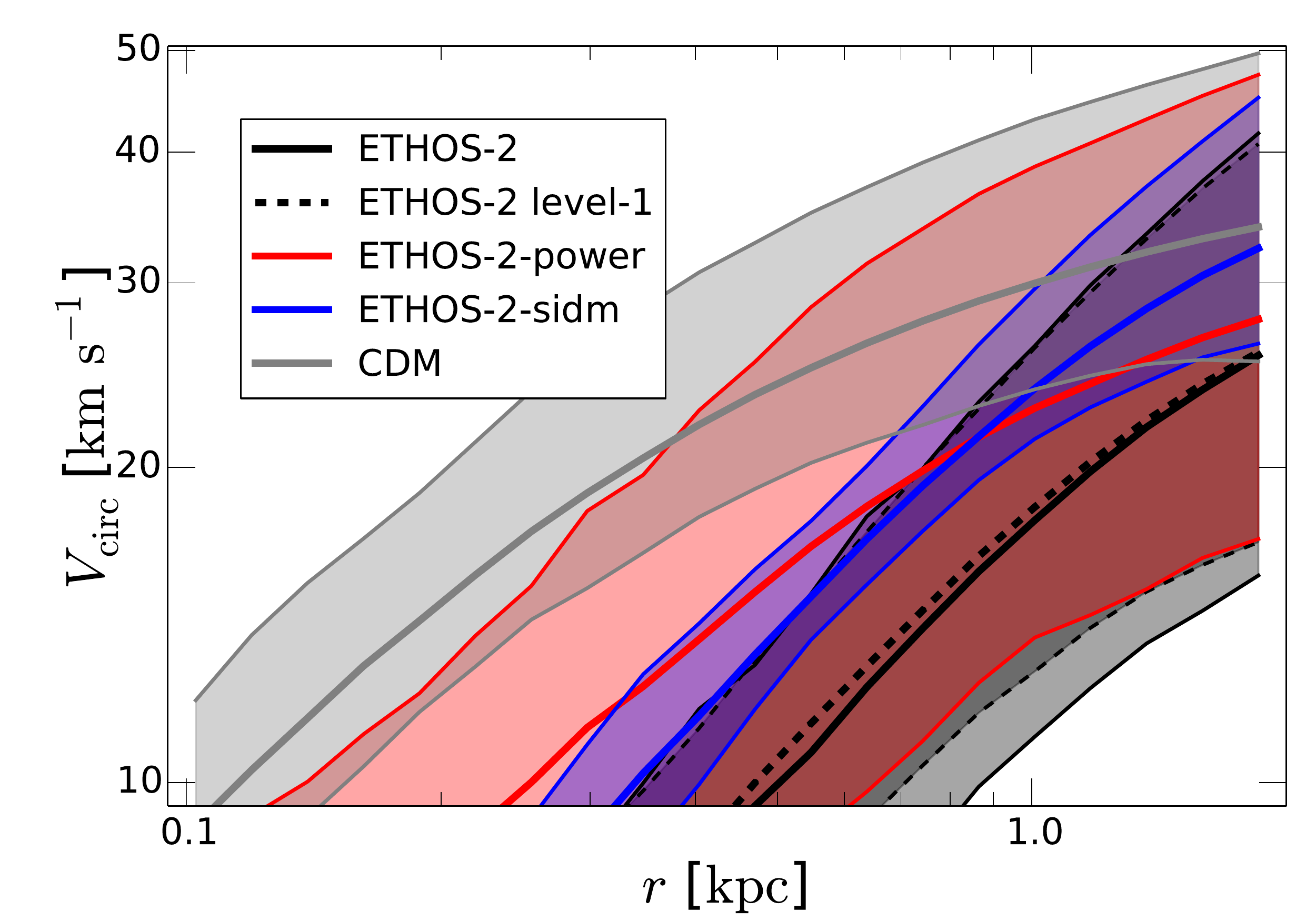}\\
\vspace{-0.5cm}
\includegraphics[width=0.475\textwidth]{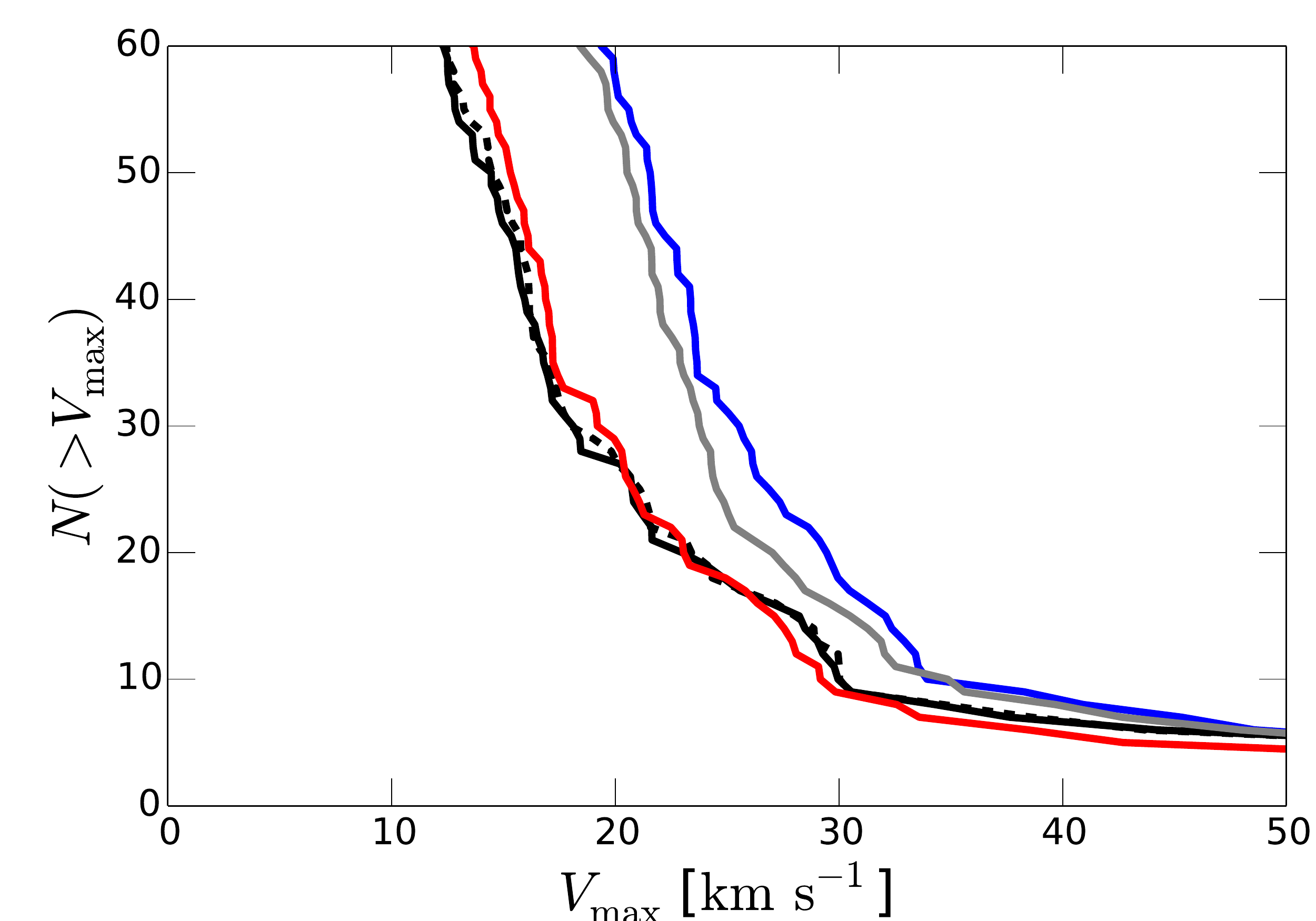}
\includegraphics[width=0.475\textwidth]{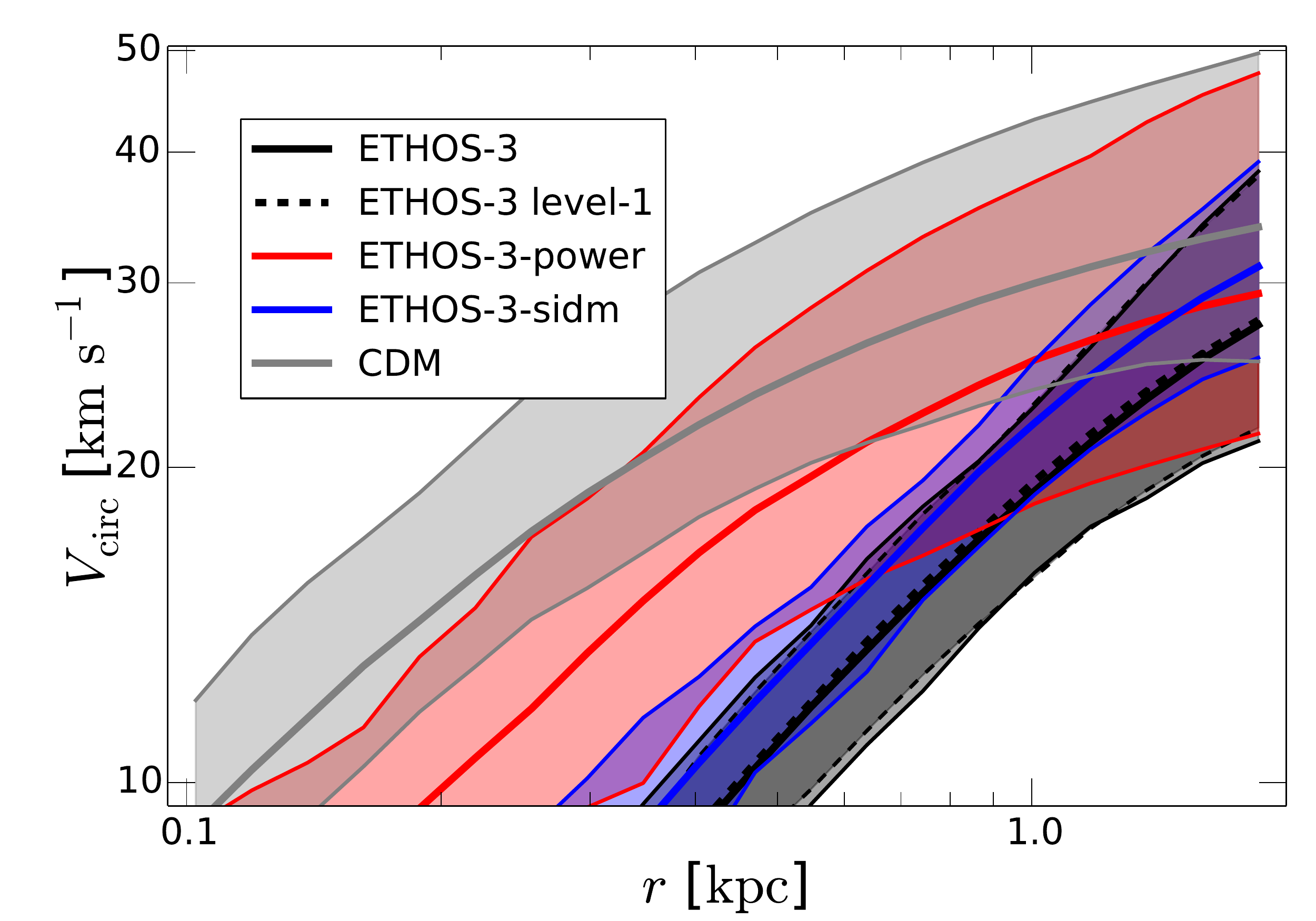}
\caption{Properties of the subhalo population at $z=0$ in the reduced DM models
summarised in Table~\ref{table:models_reduced}. Left panel: Subhalo abundance
shown as a cumulative $V_{\rm max}$ plot. Right panel: Internal subhalo
structure revealed by the circular velocity curves. We show from top to bottom
the different models ETHOS-1 to ETHOS-3 along with their reduced versions. These reduced
versions consider only DM self-interactions (ETHOS-X-sidm) or only a damping in the initial power spectrum
due to DM-DR interactions (ETHOS-X-power). The latter affects both the
abundance of subhaloes and the normalisation of the circular velocity curves.
Self-interactions only reduce the inner circular velocities (enclosed masses), but do
not alter the abundance. }
\label{fig:models_reduced}
\end{figure*}

In this section we try to disentangle the impact of the damping of the initial
power spectrum and the late-time effect of DM self-interactions. To this end we reran the
MW halo for models ETHOS-1 to ETHOS-3, at resolution level-2, which has converged reasonably in the inner structure of 
the most massive subhaloes as can be see in Fig.~\ref{fig:models_reduced}, where we compare level-1 (dashed) and level-2 (solid). For these resimulations we 
either only consider a modification to the power spectrum or only self-interactions. We can then contrast these two
variations of each model with its full version and see which new DM physics is responsible for
which effect. These different variations are summarised in Table~\ref{table:models_reduced}, and are
labelled ETHOS-X-sidm and ETHOS-X-power, where $X=1,2,3$. 

For the ETHOS-X-sidm models we only consider self-interactions and use the same 
transfer function as for the CDM simulation. The ETHOS-X-power simulations,
on the other hand, do not include self interactions, but use the corresponding
damped initial power spectra. We are specifically interested in the abundance and the internal
structure of the subhalo population at $z=0$.
In Fig.~\ref{fig:models_reduced} we show the subhalo
mass function as a function of $V_{\rm max}$ (left panel) and the circular
velocity profile for the top $15$ subhaloes (right panel) for the different reduced models
(each in a row as shown in the legends). The most striking impact can be seen when looking at the
ETHOS-X-sidm results, which do not include the damping of the initial power
spectrum at small scales. Self-interactions do not affect the abundance of subhaloes,
at least at the amplitude of the cross sections we consider here~\citep[see
also][]{Vogelsberger2012}. The values of $V_{\rm max}$ are also not affected
significantly compared to CDM, it is only within $r_{\rm max}$ where the effect
of DM collisions is clear. The larger the cross section, the stronger the reduction of
the enclosed mass within $r_{\rm max}$. We note that the dispersion in the distribution of the
subhalo densities is quite small in these reduced ETHOS-X-sidm models. This is because the 
maximum central core density cannot be lower than the density at $r_{\rm max}$, thus subhaloes with
a similar $V_{\rm max}$ and $r_{\rm max}$ will develop a similar core density. 
Modifying the initial power spectrum naturally reduces the subhalo abundance and creates a broad
dispersion in the distribution of velocity curves of the most massive subhaloes. This distribution is shifted down
towards lower densities. These properties are ultimately connected
to the (mass-dependent) delay in the formation of haloes caused by the primordial damping to the power
spectrum. Therefore a cutoff in the primordial power spectrum creates a
dispersion in the circular velocity profiles of haloes with sizes around the
cutoff scale. This might help to alleviate the problem of diversity of rotation
curves present in dwarf galaxies as pointed by~\cite{Oman2015}, albeit this problem was shown clearly only at
larger scales ($V_{\rm max}\sim100$kms$^{-1}$) than the ones studied here. Current hydrodynamical simulations fail to reproduce 
this diversity in the inner regions of dwarf galaxies; i.e. there exists currently
no viable solution for this problem within CDM even if baryonic processes are considered~\citep[][]{Oman2015}. 

From the right panel of Fig.~\ref{fig:models_reduced}, we can conclude that the
central subhalo densities in the full model ETHOS-1 are mainly due to the
damping of primordial perturbations caused by early DM-DR interactions, while
for model ETHOS-3, they are essentially given by the large amplitude of the
DM-DM collisions. Model ETHOS-2 lies somewhere in between.

\begin{table}
\begin{center}
\begin{tabular}{cc}
\hline
Name       & reduced model   \\
           &                 \\
\hline
\hline
ETHOS-1-sidm    & ETHOS-1, only self-int. w/ CDM transfer fct.\\
ETHOS-1-power   & ETHOS-1, no self-int. \\
\hline
ETHOS-2-sidm    & ETHOS-2, only self-int. w/ CDM transfer fct.\\
ETHOS-2-power   & ETHOS-2, no self-int. \\
\hline
ETHOS-3-sidm    & ETHOS-3, only self-int. w/ CDM transfer fct.\\
ETHOS-3-power   & ETHOS-3, no self-int. \\
\hline
\end{tabular}
\end{center}
\caption{Overview of reduced DM models. These models are similar to ETHOS-1 to ETHOS-3, but
they only include self-interactions without the damping of the power spectrum
(``sidm''), or they do not include self-interactions, but have the damping of
the primordial power spectrum (``power''). The reduced models help us to disentangle these two
effects present in our full models.}
\label{table:models_reduced}
\end{table}

\begin{figure*}
\centering
\includegraphics[width=0.475\textwidth]{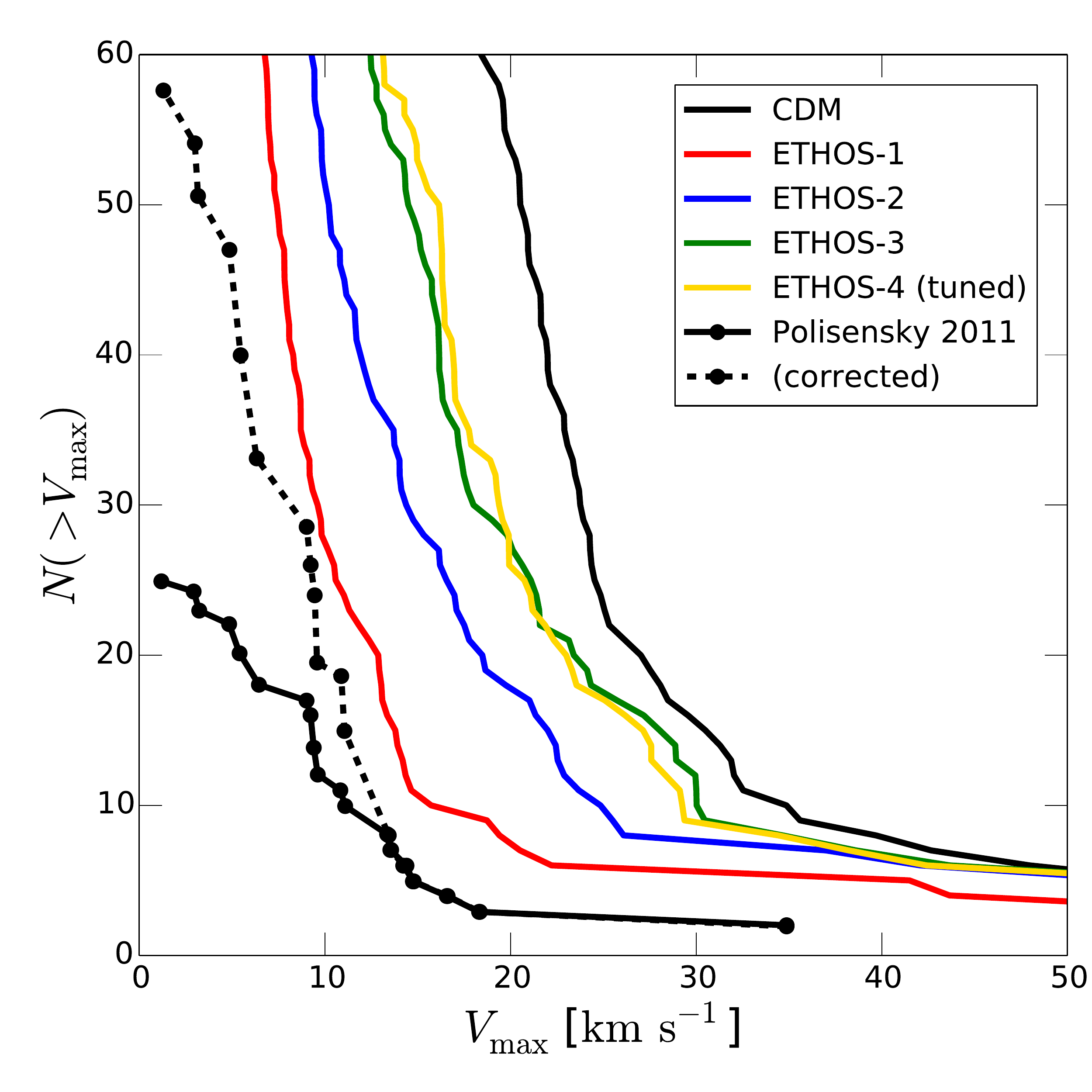}
\includegraphics[width=0.475\textwidth]{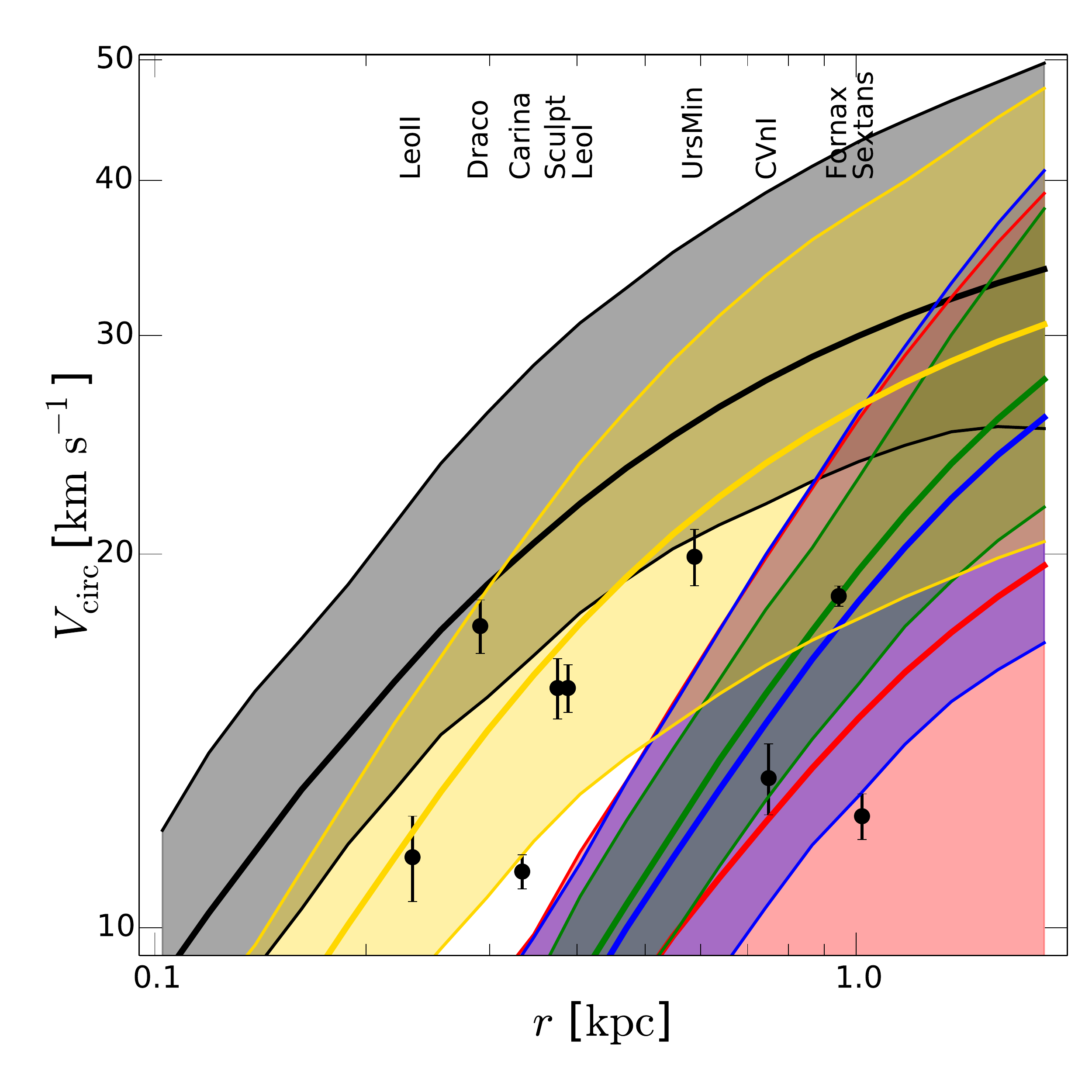}
\caption{
Subhalo population for the tuned model ETHOS-4. This model was specifically set up
to address the MS and TBTF problems. Left panel: The number of satellite
galaxies as a function of their maximal circular velocity for the four
different models with a comparison to observed satellites of the Milky Way including a sky coverage correction~\citep{Polisensky2011}. We show all subhaloes with a halocentric distance less
than $300\kpc$. Right panel: Circular velocity profiles of the same haloes. The
data points show MW dSphs taken from \protect\cite{Wolf2010}. The ETHOS-4 model provides a reasonable fit
to the subhalo population of the MW.}
\label{fig:model_m4}
\end{figure*}

\subsection{A tuned DM model}

So far we have seen that the models ETHOS-1 to ETHOS-3 affect the subhalo population of
galactic haloes in different and significant ways. However, none of these
models seem to provide a good fit to the actual satellite population of the MW.
Most importantly, these models ``over-solve'' the TBTF problem, and likely the
CC problem as well,  due to the combined effect of a strong damping of
the initial linear power spectrum and a rather large cross section at the
characteristic velocities within the satellites.  We therefore search in the
parameter space of ETHOS for a model that can alleviate the abundance and
structural problems of CDM simultaneously.  It turns out that it is actually
difficult to find a combination of damping scale and self-interaction
cross-section that alleviates both of these problems.  This difficulty implies
that by trying to solve the MW CDM problems, we need to significantly reduce
the parameter space of ETHOS.  If we assume that the solution to these problems
is largely driven by the particle nature of DM, we can therefore use this
``tuning process" to indirectly constrain our underlying DM particle
physics models. This approach is similar to that of~\cite{Boehm2014}. 

\begin{figure}
\centering
\includegraphics[width=0.475\textwidth]{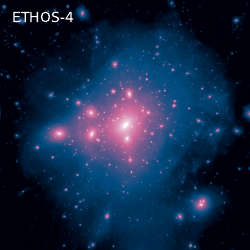}
\caption{DM density projections of the zoom MW-like halo simulations for the
tuned model ETHOS-4. The projection has a side length and depth of $500\kpc$. The
initial power spectrum is essentially the same as in ETHOS-3. The amount of
substructure and the general DM density distribution looks very similar to ETHOS-3.
Remaining differences are driven by the very different
self-scattering cross section between ETHOS-3 and ETHOS-4.}
\label{fig:dm_density_small_M4}
\end{figure}

We have used the results of models ETHOS-1 to ETHOS-3 to find a new model
(ETHOS-4) which gives a reasonable fit to the observational data of the MW
satellite population. This means that we had to iterate over different
simulations (at resolution level-2) with different cutoff scale to the power
spectrum and different cross sections, using models ETHOS-1 to ETHOS-3 as
references.  Clearly, the cutoff of this new model needed to be on smaller
scales than that of ETHOS-1 and ETHOS-2 since both models predict a very strong
impact on subhalo densities, too strong to be consistent with the kinematics of
MW dSphs.  Furthermore, the cross section of the self-interactions at the dwarf
velocity scale should be smaller than that of ETHOS-1 to ETHOS-3 since these
models (even their reduced ETHOS-X-sidm versions) over-solve the TBTF problem.
Based on these considerations we have found model ETHOS-4, as a suitable
candidate (see Table~\ref{table:models} for the parameters of this model). 
The initial
power spectra of ETHOS-3 and ETHOS-4 are essentially the same as can be seen in Fig.~\ref{fig:models} (left panel). This results in
a very similar suppression of the number of satellites, which can clearly be
seen in the left panel of Figure~\ref{fig:model_m4}. The cross section,
on the other hand, is radically different and for low velocities more than an
order of magnitude smaller than ETHOS-1. This is the key feature of model
ETHOS-4: the lower cross section results in an increased central density and
larger circular velocities compared to the ETHOS-1 to ETHOS-3 models. This can
clearly be seen in the right panel of Figure~\ref{fig:model_m4}. Model
ETHOS-4 demonstrates that we can find parameters for our particle physics model, and thus ETHOS parameters. Other particle physics models that map into similar parameters to ETHOS-4 should give similar results.
that can alleviate the main CDM tensions at small scales. We note that this
model also satisfies all current large-scale constraints as well since it
deviates even less from CDM than the models ETHOS-1 to ETHOS-3.  A density map
of the DM distribution in this model ETHOS-4 is shown in
Fig.~\ref{fig:dm_density_small_M4}.  There are only little differences with
respect to ETHOS-3, which are due to the significantly
smaller cross section of ETHOS-4 compared to ETHOS-3.  We note that despite having considerably lower cross
sections than the other models we explored, self-interactions are still relevant in ETHOS-4. We have verified that
the central densities in subhaloes are lower (albeit the effect is relatively small) in ETHOS-4 than in a setting with the same features 
but with the self-interactions turned off. 

We note that ETHOS-4 alleviates the tension between theory and
observations for the TBTF and MS problems, but our MW-size simulations cannot be used
to study directly if such a model could also produce the large cores seemingly inferred 
in low surface brightness galaxies \citep[e.g.][]{Kuzio2008}, which might require large
cross sections in an interpretation based on DM collisions~\citep[see Fig. 1 of][]{Kaplinghat2015}. However, besides pure
DM self-interactions, ETHOS-4 also includes a relevant effect due to the damping of the power
spectrum. Both effects could combine to reduce densities sufficiently to be consistent with observation of
LSB galaxies. 
Furthermore, the character of this interplay could be adjusted relative to ETHOS-4 paramaters by increasing the normalization of the cross section and increasing slightly the scale for the power spectrum cut off. This would enhance the SIDM-driven core creation, while retaining significant deviations 
from CDM in the abundance of dwarf systems. We plan to explore these issues in more detail in upcoming works.

\section{Conclusion}

We have explored simulations of self-interacting DM (SIDM), which come from a
particle physics model with a single dark matter particle interacting with
itself and with a massless neutrino-like fermion (dark radiation) via a massive mediator.
The parameters of the particle physics model are mapped into an effective
framework of structure formation (ETHOS, see \cite{Cyr-Racine2015}), where each
model is described by a specific initial power spectrum and a
velocity-dependent self-interaction cross section.

In this paper we have analysed a few benchmark cases in the large parameter
space of ETHOS, which mainly have relevant consequences for the formation and
evolution of dwarf-scale haloes.  We have simulated these cases in
$(100\hmpc)^3$ uniform boxes with $1024^3$ DM particles.  A galactic
Milky-Way-size halo was then selected from this box and simulated at higher
resolution.  Our main focus is the study of this galactic halo and its subhalo
population. Our highest resolution simulation has a Plummer-equivalent
softening length of $72.4\pc$, and a mass resolution of $2.8\times 10^4\msun$.  We
simulate this halo in five different models: CDM and four SIDM models (ETHOS-1 to ETHOS-4).
Our main conclusion is that such models can not only change the internal
structure of subhaloes, but also affect the subhalo abundance in a similar way
to warm dark matter (WDM) models, due to the inherit damping in the initial
power spectrum.  However, unlike WDM models, the damping in our effective models
is much richer since it also contains oscillatory features caused by
interaction between dark matter and dark radiation in the early Universe.

\begin{figure}
\centering
\includegraphics[width=0.23\textwidth]{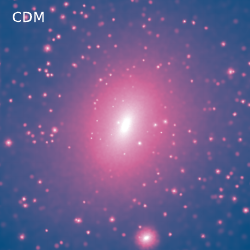}
\includegraphics[width=0.23\textwidth]{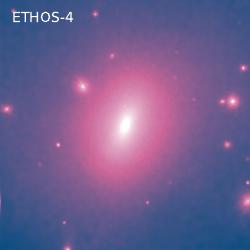}\\
\hspace{0.075cm}\includegraphics[width=0.23\textwidth]{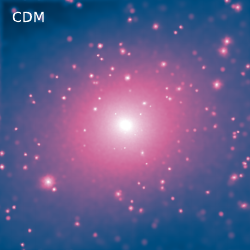}
\includegraphics[width=0.23\textwidth]{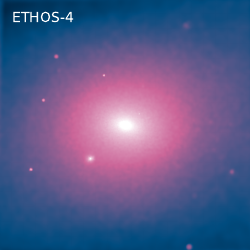}
\caption{DM density projections of the two most massive subhaloes (within
$300\kpc$ halocentric distance) in the MW-like halo simulations for CDM and the
tuned model ETHOS-4. The projection has a side length and depth of $50\kpc$. }
\label{fig:dm_density_subhalo}
\end{figure}

Our main findings are:

\begin{itemize}
\item The large scale structure is unaffected in all our non-CDM models. At
$z=0$, the matter power spectra of the different models agree with that of CDM
for $k \lesssim 200\hmpc$ (Fig.~\ref{fig:power}). The halo mass functions also
agree above $\sim 10^{11}\hmsun$, but there is a clear departure from CDM below
this scale, which is mostly driven by the primordial damping in the power
spectrum (see Fig.~\ref{fig:massfct}). We complement our simulations by
analytical insight and provide a mapping from the primordial damping scale in
the power spectrum to the cutoff scale in the halo mass function and the
kinetic decoupling temperature.

\item The inner (core) density (within a fixed physical radius of $8.7\kpc$) is
reduced mostly in the low mass haloes since the impact of self-interactions and
power spectrum damping is the largest at those masses in the models we
explored. Inner densities are affected below $10^{12}\hmsun$ and can be reduced
by up to $30\%$ for haloes around $10^{10}\hmsun$, relative to CDM
(Fig.~\ref{fig:density_profiles}). In the model with the largest cross section
(ETHOS-3), we also find a mild reduction of the central density in cluster scale
haloes.

\item The density profile of MW-size haloes shows a small core $\lesssim 2\kpc$
in the SIDM models, with the core size being the largest for the model with the
largest scattering cross section (Fig.~\ref{fig:main_profile}).

\item The subhalo abundance is strongly affected by the damping of the initial
linear power spectrum. The selected models span the whole range between the CDM
prediction and the observed satellite population (completeness corrected) of
the Milky Way (Fig.~\ref{fig:sub_vmax}). One of our models, ETHOS-1, would most
likely be ruled out by observational data if baryonic processes were to be
included, i.e. supernovae feedback, early heating by reionisation and tidal
stripping.

\item The internal structure of subhaloes is affected by both self-interactions
and the primordial damping of the power spectrum reducing the enclosed mass in
the inner regions and  producing central density cores for the most massive
subhaloes. Three of our benchmark cases (ETHOS-1 to ETHOS-3) ``over-solve'' the
too-big-to-fail (TBTF) problem in the sense that they reduce the central mass of
subhaloes too strongly. The resulting circular velocity curves then lie below
the observational data points coming from the inferred kinematics of the
classical MW dSphs (Fig.~\ref{fig:sub_internal}). This implies that ETHOS models
can actually be constrained by comparing to observational data. The large impact on
the structure and appearance of massive subhaloes can also be seen in Fig.~\ref{fig:dm_density_subhalo}, where we show density
maps of the two most massive subhaloes for the CDM and ETHOS-4 model.

\item We have searched over the parameter space of ETHOS to construct one model (ETHOS-4),
which solves the TBTF problem and at the same time alleviates the
missing satellite (MS) problem (Fig.~\ref{fig:model_m4}). 

\item We also notice that introducing a cutoff in the primordial power spectrum
(in our case caused by DM-DR interactions), is a natural way to create a
dispersion in the circular velocity profiles of haloes with sizes around the
cutoff scale. This might help to alleviate the problem of diversity of rotation
curves present in dwarf galaxies~\citep[][]{Oman2015}; albeit this problem has only been reported at scales
larger than the ones discussed here.  We stress that
current hydrodynamical simulations fail to reproduce this diversity in the inner regions of dwarf galaxies; i.e., there exists currently
no viable solution for this problem within CDM even if baryonic processes are considered. 

\end{itemize}

We have demonstrated that despite the larger accessible parameter space of our
particle physics models, it is by no means trivial to find a viable and
promising DM solution to some of the small-scale problems of galaxy formation.
Instead, we found a surprising non-linear amplification of the effects of late
DM self-interactions and early DM-DR interactions on the inner velocity
profiles, which complicates any simple inference of the underlying particle
physics nature from astrophysical constraints.  We conclude that the effective
models discussed here provide an interesting alternative to CDM by being
capable of solving some of its outstanding small scale problems by exploring a
richer (but allowed) dark matter phenomenology.  Specifically, those models
might be able to solve or alleviate the too-big-to-fail problem, the missing
satellite problem, the core-cusp problem and produce a larger diversity of dwarf
galaxy rotation curves simultaneously.

The parameter space of this generalisation of the structure formation theory
can be constrained by contrasting model predictions to observations,
particularly when baryonic processes are added into the theory. Albeit the
increased parameter space makes this effective framework more complex than CDM, it
is a complexity that is necessary given our incomplete knowledge of both the
dark matter nature and the strength of the key baryonic processes affecting
galaxy formation and evolution. Clearly, our effective theory needs to be explored in much more detail including its interplay 
with baryonic physics. We will address these questions in the future.

\section*{Acknowledgements}
We thank Michael Boylan-Kolchin, Federico Marinacci for useful comments, and Volker Springel for giving us access to {\tt AREPO}.  The simulations were
performed on the joint MIT-Harvard computing cluster supported by MKI and FAS.
MV acknowledges support through an MIT RSC award.  The Dark Cosmology Centre is
funded by the DNRF. JZ is supported by the EU under a Marie Curie International
Incoming Fellowship, contract PIIF-GA-2013-62772. KS gratefully acknowledges support from the Friends of the Institute for Advanced Study.  The research of KS is supported in part by a National Science and Engineering Research Council (NSERC) of Canada Discovery Grant. F.-Y. C.-R. acknowledges the support of the National Aeronautical and Space Administration ATP grant 14-ATP14-
0018 at Harvard University. The work of F.-Y. C.-R. was performed in part at the California Institute of Technology
for the Keck Institute for Space Studies, which is funded by the W. M. Keck Foundation. C.P. gratefully acknowledges the support of the Klaus Tschira Foundation.


\label{lastpage}

\end{document}